\documentclass[aps, prb, superscriptaddress, twocolumn, letterpaper, 10pt, amsfonts, amsmath, amssymb]{revtex4-2}

\usepackage{blindtext}
\usepackage{graphicx}
\usepackage{subfigure}
\usepackage{hyperref} 
\hypersetup{
    colorlinks,
    linkcolor={blue!80!black},
    citecolor={blue!80!black},
    urlcolor={blue!80!black}
}

\usepackage{xcolor}
\usepackage{tikz}
\usepackage{pgfplots}
\pgfplotsset{
colormap name=viridis,
compat=1.16
}
\usetikzlibrary{decorations.pathreplacing}
\usepackage{soul}

\usepackage{physics}

\usepackage{placeins}

\usepackage{siunitx}
\usepackage{chemformula}

\usepackage{xcolor}

\definecolor{AMColor}{rgb}{0.16, 0.23, 0.55}

\definecolor{editcolor}{rgb}{1, 0.0, 0.0}

\begin{document}
\title{Demonstrating Majorana non-Abelian properties using fast adiabatic charge-transfer}

\author{Svend Krøjer}
\affiliation{Center for Quantum Devices, Niels Bohr Institute, University of Copenhagen, DK-2100 Copenhagen, Denmark}
\author{Rubén Seoane Souto}
\affiliation{Center for Quantum Devices, Niels Bohr Institute, University of Copenhagen, DK-2100 Copenhagen, Denmark}
\affiliation{Division of Solid State Physics and NanoLund, Lund University, S-22100 Lund, Sweden}
\author{Karsten Flensberg}
\affiliation{Center for Quantum Devices, Niels Bohr Institute, University of Copenhagen, DK-2100 Copenhagen, Denmark}

\date{\today}

\begin{abstract}
\noindent Demonstration of Majorana non-Abelian properties is a major challenge in the field of topological superconductivity. In this work, we propose a minimal device and protocol for testing non-Abelian properties using charge-transfer operations between a quantum dot and two Majorana bound states combined with reading the parity state using a second dot. We use an adiabatic perturbation theory to find fast adiabatic paths to perform operations and to account for nonadiabatic errors. We find the ideal parameter sweep and a region in parameter space which reduces the charge-transfer operation time 1-2 orders of magnitude with respect to constant velocity driving. Using realistic parameters, we estimate that the lower bound for the time scale can be reduced to $\sim10$ ns. Deviations from the ideal parameters lead to the accumulation of an undesired dynamical phase, affecting the outcome of the proposed protocol. We furthermore suggest to  reduce the influence from the dynamical phase using a flux echo. The echo protocol is based on the $4\pi$-periodicity of the topological state, absent for trivial bound states.

\end{abstract}

\maketitle

\section{Introduction}

The realization and verification of Majorana bound states (MBSs) have received a substantial amount of attention in the past decade \cite{NayakReview,LeijnseReview,AguadoReview,LutchynReview,BeenakkerReview_20}.
MBSs are exotic zero-energy quasiparticle states appearing at the ends of one-dimensional topological superconductors (TSs) or in vortices of two-dimensional TSs \cite{Fu2008,Lutchyn2010,Oreg2010}. 
\begin{figure}[ht]
    \centering
    \includegraphics[width=0.8\columnwidth]{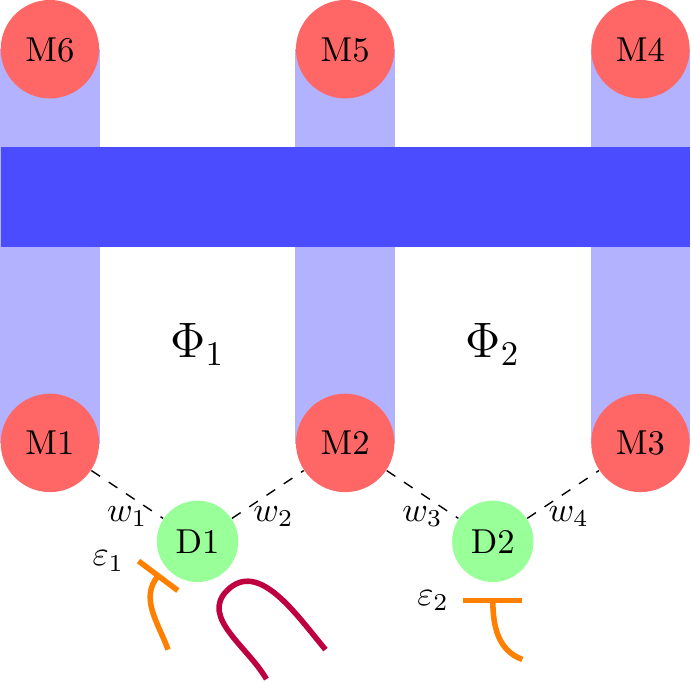}
    \caption{Schematic of the proposed device for demonstrating MBSs non-Abelian properties. Three long TS nanowires (light blue) extend from a trivial superconducting backbone (blue). MBSs (red) form at the ends of the TSs. M1, M2 and M3 are tunnel coupled to quantum dots (green) D1 and D2 with coupling strengths $w_i$. The dot energies $\varepsilon_i$ are controlled with nearby gates (orange). In our protocols, D1 is used for initialization and read out of the M1/M2 pair using a charge sensor (purple). D2 is used for charge-transfer processes involving the M2/M3 pair \cite{Flensberg2011,Souto2020}. 
    Magnetic fluxes $\Phi_1, \Phi_2$ control the splitting between the even and odd parity states. The remaining MBSs (M4, M5 and M6) are separated from M1, M2 and M3 and do not contribute to the system dynamics.}
    \label{fig:device}
\end{figure}
MBSs exhibit non-Abelian exchange properties contrary to topologically trivial subgap states \cite{Ivanov2001,Alicea2011}. 
Experimental demonstration of MBSs non-Abelian properties is one of the key goals in the field as it will probe their topological origin, distinguishing them from trivial states. An additional promising feature of MBSs is their ability to store quantum information in non-local fermionic degrees of freedom, becoming robust to local perturbations \cite{AguadoReview}. In this way, MBSs can encode quantum information in the degenerate ground-state manifold. Braiding operations (exchange of MBSs) can perform Clifford gates, thus implementing (non-universal) topological quantum computing \cite{NayakReview}.

To experimentally realize MBSs, a number of structures and devices have been proposed \cite{FlensbergReview2021}. Hybrid semiconductor-superconductor heterostructures are widely used platforms in the attempt to realize one-dimensional spin-polarized $p$-wave superconductors hosting MBSs at its ends \cite{Lutchyn2010,Oreg2010}. 
Recent progress on fabrication techniques has made it possible to measure signatures consistent with MBSs. Early observations include the measurement of a robust zero-bias conductance peak \cite{Mourik2012,Deng2016}. Later experiments indicated the $ 2e^2/h $-quantization of the zero-bias peak \cite{Nichele2017}. Measurements have shown coherent transport through a Majorana island \cite{Whiticar2020}, exponential scaling of energy separation with length \cite{Albrecht2016,Vaitiekenas2020}, and hybridization characteristics with quantum dot states \cite{Deng2016,Deng2018}.
Despite the mounting signatures consistent with MBSs, direct observation of their non-Abelian exchange properties remains a challenge in the field. Such demonstration could provide smoking-gun evidence for the topological origin of MBSs, while having the outlook of being a first step in implementing protected gates in Majorana qubit devices.

In practice, showing non-Abelian exchange properties through real space braiding of MBSs in T- or Y-junctions is expected to be a great experimental challenge as it is difficult to tune in and out of the topological regime \cite{Alicea2011,Harper_PRR2019}. For this reason, this paper instead focuses on implementing braiding-like operations of MBSs in parameter space. Following Refs. \cite{Flensberg2011,Souto2020}, we consider manipulating the occupation of MBSs through charge-transfer processes with a nearby quantum dot in the Coulomb-blockaded regime, see Fig.\ \ref{fig:device} for a device schematic similar to Ref.\ \cite{Flensberg2011}. In a successful charge-transfer process, an electron is adiabatically exchanged between the gate-controlled quantum dot and the MBSs, changing the Majorana parity. An advantage of this parameter space operation is that it generalizes the real space braiding to rotations through a continuum of angles, extending the space of possible operations through braiding operations alone. The immediate downside, however, is that charge-transfer operations are not topologically protected and require accurate tuning of the parameters to achieve high fidelity.

noncommutativity of braiding-like operations can provide evidence for the non-Abelian nature of MBSs.
Concretely, we search for protocols where interchanging two charge-transfer operations influence the measured parity of the Majorana state. A protocol consists of two sequences with charge-transfer operations applied in different order, testing the noncommutativity of the operations. \cite{Flensberg2011}.
In the device shown in Fig.\ \ref{fig:device}, the principal source of error is due to splitting of the ground state degeneracy with imperfect tuning of the parameters. This leads to a relative dynamical phase between the split states, reducing the visibility of the geometric phase originated from non-Abelian charge-transfer operations.
As the charge-transfer process is meant to operate on long, adiabatic time scales, even a small energy splitting can lead to a substantial relative phase error, overwhelming the non-Abelian signal. This presents a trade-off between driving the system slowly enough to remain in the ground state and fast enough to avoid the effects of the splitting.

\begin{figure}[thp]
    \centering
        \includegraphics[width=0.9\columnwidth]{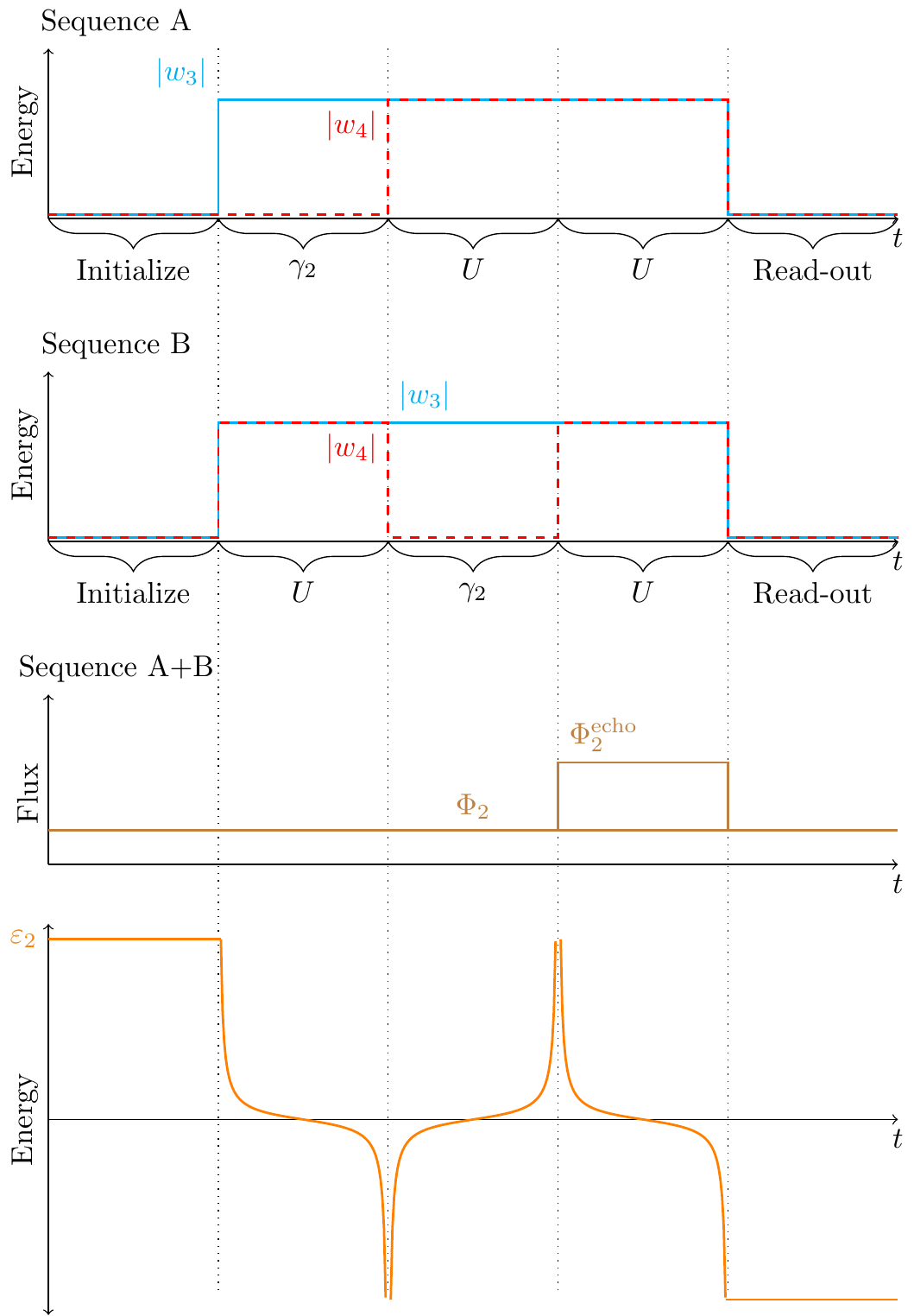}
    \caption{Diagram of the two sequences (top and middle panels), consisting of three charge-transfer processes. In each diagram, both the protocols with and without the echo effect are depicted. \textbf{Top}: Sequence A. Here, $|w_4|$ is initially set to zero during the first charge-transfer process. For the two subsequent charge-transfer processes it is ideally set to $|w_4|=|w_3|$. As indicated, the echo protocol is achieved by adjusting the magnetic field before the third charge-transfer process. \textbf{Middle:} Sequence B. Here, $|w_4|$ is instead set to zero during the second charge-transfer process, reversing the order of the first two operations. \textbf{Bottom:} Level energy of D2 for both sequences.}
    \label{fig:protocol}
\end{figure}

In this study, we propose an experiment for testing the non-Abelian properties of MBSs. We simplify the device and reduce the number of operations needed with respect to the original proposal in Ref.\ \cite{Flensberg2011}. We improve the visibility of the MBSs non-Abelian signature by optimizing the adiabatic charge-transfer processes. We also design a $4\pi$-periodic flux echo protocol that cancels the undesired dynamical phase of subsequent operations.

Specifically, our device and protocol proposals are minimal as they require controlling a single quantum dot (D2) and one tunneling amplitude ($w_{4}$), see Fig. \ref{fig:device}. A second quantum dot, D1, is used to measure the parity of the non-local fermion formed by M1 and M2 \cite{Munk_PRR2020,Steiner_PRR2020,Smith2020, Szechenyi2020,Khindanov2021,Schulenborg2021}. We propose two variants of the protocol: with and without the echo mechanism. Both protocols, depicted in Fig.\ \ref{fig:protocol},
require using one dot and three adiabatic charge-transfer processes. In the flux echo protocol, the dynamical phase is canceled by flipping the sign of the energy splitting in between charge-transfer operations. This is accomplished by tuning the magnetic field $\Phi_2$ to induce an additional superconducting (SC) phase difference, flipping the sign of the energy splitting between the even and odd parity ground states.
We find that the echo protocol is robust to drifts in the SC phase difference and that the deviations in the additional SC phase can be as large as $\sim$10\% from the ideal value, $2\pi$. As the flux echo relies on the $4\pi$-periodicity of Majorana parity states, it also makes it possible to distinguish from $2\pi$-periodic trivial states. 

To mitigate nonadiabatic and phase errors, we formulate a consistent theoretical framework for finding fast, adiabatic paths based on adiabatic perturbation theory (APT) developed in Ref.\ \cite{Rigolin2008}. Within the framework, we find how to optimally control the level energy of the quantum dot to minimize the dynamical phase without introducing nonadiabatic errors such as transitions to excited states. Compared to driving the system with constant (Landau-Zener) velocity, we find an adiabatic path that is one to two orders of magnitude faster than a linear sweep of D2 energy, as used in Ref. \cite{Souto2020}. We provide numerical calculations supporting these results. Finding fast adiabatic paths is crucial for adiabatic quantum computing as discussed by previous attempts \cite{Roland2001,Schaller2006,Rezakhani2009,Martinis2014}. Specifically in the context of Majorana-based systems, the velocity of real space exchange and operations using varying tunnel couplings between MBSs has been considered \cite{Cheng2011,Scheurer2013,Karzig2013,Karzig2015, Karzig2015Shortcuts,Knapp2016,Rahmani2017,Sekenia2017,Ritland2018,Nag2019,Zhang2019,Posske2020,Breckwoldt2021}. In this work, we instead consider the nonadiabatic effects that occur when MBSs are coupled to a driven quantum dot.

\section{Theory}\label{sec:model}

We begin by reviewing the charge-transfer process following Ref.\ \cite{Flensberg2011} and formulate the non-Abelian operations in terms of the relative geometric phase between the even and the odd parity ground states. This enables us to identify the non-Abelian operations resulting from charge-transfer processes where the ground states energy split.

Then, we review the adiabatic perturbation theory following Ref.\ \cite{Rigolin2008} and formulate a framework for studying fast adiabatic processes, resulting in predictions for the optimal charge control. 

\subsection{Charge-transfer process}\label{sec:charge-transfer-process}

To describe the charge-transfer process between the quantum dot D2 and the MBSs M2 and M3 (see Fig.\ \ref{fig:device}), we consider the low-energy Hamiltonian \cite{Flensberg2011},
\begin{equation}\label{eq:HamD2}
    H=\varepsilon_2 d_2^\dagger d_2+(w_3^* d_2^\dagger-w_3 d_2)\gamma_2+(w_4^* d_2^\dagger-w_4 d_2)\gamma_3.
\end{equation}
The first term describes D2 with $\varepsilon_{2}$ being its time-dependent energy and $d_{2}$ its electron annihilation operator. The second and third terms in Eq.\ \eqref{eq:HamD2} describe the tunnel coupling to M2 and M3, with $w_3$ and $w_4$ being the tunneling amplitudes. Here, $\gamma_2$ and $\gamma_3$ are the self-adjoint Majorana operators.


Our proposed protocol is based on operating on the state of M23 using D2.
The annihilation operator of the M23 fermion is defined by $ f_{23}=1/2(\gamma_2+i\gamma_3)$, giving a Hilbert space of dimension four. Due to the total parity conservation, the Hamiltonian matrix corresponding to Eq.\ \eqref{eq:HamD2} is block diagonal with even and odd parity blocks given by,
\begin{equation}\label{eq:HamMatrix}
    \mathcal{H^\rho}=\begin{pmatrix}
    0 & w^\rho \\
    (w^\rho)^* & \varepsilon_2
    \end{pmatrix},
\end{equation}
where $w^\rho=w_3-\rho\, i\, w_4$. We use the even basis ($\rho=+$) $\{ \ket{0}_\text{D2}\ket{0}_\text{M23},\ket{1}_\text{D2}\ket{1}_\text{M23} \}$ and odd basis ($\rho=-$) $\{ \ket{0}_\text{D2}\ket{1}_\text{M23},\ket{1}_\text{D2}\ket{0}_\text{M23} \}$, with $0(1)$ referring to the occupation of D2 and M23. 
 
We parametrize the tunnel couplings as $w_3=w\, e^{i\phi/2}\cos\theta$ and $ w_4=w\sin\theta$ where the magnetic flux $\Phi_2$ controls the SC phase difference $\phi=\Phi_2/(h/(2e))$. Here, $\theta$ controls the asymmetry on the tunnel coupling strength. The eigenenergies of the Hamiltonian matrix in Eq.\ \eqref{eq:HamMatrix} are
\begin{equation}\label{eq:eigenenergy}
    E_\pm^\rho=\varepsilon_2/2\pm\sqrt{(\varepsilon_2/2)^2+w^2(1-\rho\sin(2\theta)\sin(\phi/2))},
\end{equation}
with the corresponding eigenstates
\begin{equation}\label{eq:eigenstates}
    \psi_\pm^\rho=\frac{1}{\sqrt{(E_\pm^\rho)^2+|w^\rho|^2}}\begin{pmatrix}
    w^\rho \\
    E_\pm^\rho
    \end{pmatrix}.
\end{equation}
The energy spectrum of the system is $4\pi$-periodic, and the even and the odd parity sectors are degenerate at integer values of $\phi/(2\pi)$.

In a successful charge-transfer process, an electron is transferred between D2 and the fermion formed by M23. This is accomplished by inverting the energy on D2 from $\varepsilon_0$ to $-\varepsilon_0$, allowing the exchange of a charge.
The initial and final level energies are not required to be equal in magnitude but they should be much larger than the coupling strength to D2. We assume $\varepsilon_0>0$ in what follows and disregard the effect from the continuum of states by taking the limit $\Delta_{\text{SC}}>\varepsilon_0\gg w$. The effect of the continuum of states above the superconducting gap $\Delta_{\text{SC}}$ has been discussed in Ref.\ \cite{Souto2020}. We further assume that the time $T$ of the charge-transfer process is shorter than the quasiparticle poisoning time scale, yet long enough for the process to be adiabatic.

To understand the non-ideal charge-transfer operations, it is helpful to consider the geometric phase acquired by the even parity ground state relative to the odd parity ground state. 
Since the charge-transfer process is not a loop in parameter space, the calculation of the geometric phase is slightly subtle and can be found in Appendix \ref{sec:appendix_1l}. The accrued relative geometric phase between the even and odd parity ground states during a process where the dot is filled ($\varepsilon_2: \varepsilon_0\to-\varepsilon_0,\, \varepsilon_0>0$) is
\begin{equation}\label{eq:GeomPhase}
    \theta^G=\arctan[\tan(2\theta)\cos(\phi/2)],
\end{equation}
with corrections of order $(w/\varepsilon_0)^2$.
The corresponding operation on the MBSs is
\begin{align}\nonumber
    U^G&=e^{i\theta^G/2} f_{23}^\dagger+e^{-i\theta^G/2} f_{23}\\
    &=\cos(\theta^G/2)\gamma_2+\sin(\theta^G/2)\gamma_3.
    \label{eq:GeomOperation}
\end{align}
When the dot is filled, an electron tunnels from the superconductor to the dot. In the odd parity sector, the electron jumps from the occupied M23 fermionic state ($d_2^\dagger f_{23}$). In the even sector, the M23 fermion state is vacant. In this case, a Cooper pair splits with one electron occupying the M23 state while the other tunnels to the dot ($d_2^\dagger f_{23}^\dagger$).
Isolating the part acting on the M23 fermion and inserting the relative geometric phase, we arrive at Eq.\ \eqref{eq:GeomOperation}.
For the reverse process, the sign of the geometric phase and the roles of even and odd sectors with regards to the tunneling are both interchanged. For this reason, Eq.\ \eqref{eq:GeomOperation} also holds when emptying the dot. In the ideal situation, integer $\phi/(2\pi)$ and adiabatic dot energy sweep, our result simplifies to $U^G=\cos\theta\, \gamma_2+\sin\theta\, \gamma_3$, agreeing with the original result found in Ref. \cite{Flensberg2011}.

It is straightforward to relate the relative phase between the even and odd ground states to a parity-measurement of the fermion formed by the M12 pair using the dot D1. In the measurement-basis, we define $f_{12}=1/2(\gamma_1+i\gamma_2)$ and $f_{34}=1/2(\gamma_3+i\gamma_4)$ with even $\{\ket{0}_\text{M12}\ket{0}_\text{M34}, \ket{1}_\text{M12}\ket{1}_\text{M34}\}$ and odd $\{\ket{0}_\text{M12}\ket{1}_\text{M34}, \ket{1}_\text{M12}\ket{0}_\text{M34}\}$ occupation states. We take as an example the ideal situation where $\theta^G/2=\theta$. Our proposed device can only initialize the fermion M12 so we consider the initial state $\ket{0}_\text{M12}\ket{\psi}_\text{M34}$ where $\ket{\psi}_\text{M34}=\alpha\ket{0}_\text{M34}+\beta\ket{1}_\text{M34}$ is a ground state. The final state after the charge-transfer operation is found by applying $U^G$ to the initial state,
    \begin{align}\nonumber
        U^G\ket{0}_\text{M12}\ket{\psi}_\text{M34}
        =i&\cos\theta\ket{1}_\text{M12}\ket{\psi}_\text{M34}\\ +&\sin\theta\ket{0}_\text{M12}\ket{\psi'}_\text{M34},\label{eq:AngleToReadout}
    \end{align}
where $\ket{\psi'}_\text{M34}=\alpha\ket{1}_\text{M34}+\beta\ket{0}_\text{M34}$.
Using the dot D1 to measure the occupation of the M12 fermion gives the result $f_{12}^\dagger f_{12}=0(1)$ with probability $\sin^2\theta$ $(\cos^2\theta)$, which does not depend on the initial state of the M34 pair. In this way, the relative phase between the even and odd ground states could be experimentally inferred from statistics.

Away from the degeneracy point, integer $\phi/(2\pi)$, the even- and odd-parity ground-states also acquire a relative dynamical phase, $\theta^D$, affecting the outcome of the final measurement. 
In Sec.\ \ref{sec:applyAPT}, we compute the relative dynamical phase for the charge-transfer process we consider, see Eq. \eqref{eq:DynPhase}. The relative dynamical phase, unlike its geometric counterpart, does not switch sign when reversing the charge-transfer process and its contribution accumulates with successive processes. This makes a difference in the operations on the MBSs when filling or emptying the dot. Including the relative dynamical phase to Eq.\ \eqref{eq:GeomOperation}, the operation depends on whether the dot is emptied ($-$) or filled ($+$),
\begin{align}\nonumber
    U&=e^{i(\theta^G\mp\theta^D)/2}f_{23}^\dagger+e^{-i(\theta^G\mp\theta^D)/2}f_{23}\\ \label{eq:fulloperator}
    &=\cos(\frac{\theta^G\mp\theta^D}{2})\gamma_2+\sin(\frac{\theta^G\mp\theta^D}{2})\gamma_3.
\end{align}
This is the full operator acting on the ground state of the system after a charge-transfer process away from the degeneracy point. The relative geometric and dynamical phases $\theta^G$ and $\theta^D$ are given in Eqs.\ \eqref{eq:GeomPhase} and \eqref{eq:DynPhase}.

\subsection{Protocol}\label{sec:protocol}

A charge-transfer operation changes the parity of the superconductor regardless of whether it is in its trivial or topological phase. It is therefore insufficient to perform only a single operation to  distinguish between topologically trivial and nontrivial subgap states. To probe the non-Abelian properties associated with topologically nontrivial states, we instead test the noncommutativity of operations executed on the degenerate Majorana subspace. In our proposed experiment, we compare the resulting states after executing two sequences of operations. These sequences consist of the same set of operations ordered in different ways, see Fig.\ \ref{fig:protocol}. The dot D1 is used to initialize and measure the occupation of the M12 Majorana pair. Applying the two sequences on the initial state $\ket{0}_\text{M12}\ket{\psi}_\text{M34}$ give the following final states,

\noindent Sequence A:
\begin{align}\nonumber
U\, U\, \gamma_2\ket{0}_\text{M12}\ket{\psi}_\text{M34}=i&\cos\theta^D\ket{1}_\text{M12}\ket{\psi}_\text{M34}\\
    +&\sin\theta^D\ket{0}_\text{M12}\ket{\psi'}_\text{M34}.\label{eq:seqA}
\end{align}
Sequence B:
\begin{align}\nonumber
    U\, \gamma_2\, U\ket{0}_\text{M12}\ket{\psi}_\text{M34}=
    i&\cos(\theta^G+\theta^D)\ket{1}_\text{M12}\ket{\psi}_\text{M34}\\
    +&\sin(\theta^G+\theta^D)\ket{0}_\text{M12}\ket{\psi'}_\text{M34}.\label{eq:seqB}
\end{align}
Here, we assume that the energy sweeps during the charge-transfer processes are adiabatic. We also take the parameters $\theta$ and $\phi$ to be the same for the operations $U$. The operation $\gamma_2$ performs a charge-transfer process where $w_4$ is turned off (corresponding to $\theta=0$), without inducing any relative phase between the even and odd parity sectors. The order of the first two operations in Eqs. \eqref{eq:seqA} and \eqref{eq:seqB} is switched between sequence A and B. Due to the noncommutativity of $\gamma_2$ and $U$, each sequence has a different geometric phase.
This difference can be sampled statistically by measuring the occupation of the M12 Majorana pair using the dot D1 \cite{Munk_PRR2020}. In the final measurement, the probability of measuring the state $\ket{0}_\text{M12}$ is $\sin^2(\theta^D)$ and $\sin^2(\theta^G+\theta^D)$ for the sequences A and B. 
In the ideal situation, integer $\phi/(2\pi)$, the relative phases simplify to $\theta^D=0$ and $\theta^G=2\theta$. The two sequences are maximally distinguishable for $\theta=\pi/4$, corresponding to symmetric coupling $w_3=w_4$. For these finely tuned values, the final state is $\ket{1}_\text{M12}$ and $\ket{0}_\text{M12}$ for the sequences A and B.

The dynamical phase, $\theta^D$, acquired during the operations described in Eqs.\ (\ref{eq:seqA}, \ref{eq:seqB}) can overwhelm the Majorana signature, coming from $\theta^G$.
This effect of $\theta^D$ can be mitigated using a mechanism similar to the spin-echo used in spin qubits \cite{Jones2000}. In Majorana devices, parity echo or flux echo have been proposed to increase the fidelity of certain operations \cite{Karzig2016,Liu2021Arxiv}. We consider implementing a flux echo based on the following observation: the relative dynamical phase in Eq.\ \eqref{eq:DynPhase} depends on the SC phase difference as $\theta^D\propto\sin(\phi/2)$. Due to the $4\pi$-periodicity, changing $\phi\to\phi+2\pi$, the sign of $\theta^D$ changes. In this way, the dynamical phase contributions from subsequent operations cancel out. Concretely, we propose to adjust the SC phase difference by tuning the magnetic flux $\Phi_2$ and set its value to $\phi$ when performing the first two charge-transfer process in Fig.\ \ref{fig:protocol}. Ideally, $\phi/(2\pi)$ is integer, but presumably it is difficult to assess its value in experiment and it may drift. Then, for the last operation, the SC phase difference is tuned $\phi\to\phi+2\pi$. Optimally, this cancels the dynamical phase in the two $U$ operations in sequences A and B. This is contrasted by trivial states whose $2\pi$-periodic spectrum will not see the effect of the flux echo. 

An advantage of this flux echo is that the required change in the SC phase difference is independent of the (unknown) value of $\phi$. This is in contrast to proposals such as $\phi\to-\phi$ which also flips the sign of the relative dynamical phase \cite{Liu2021Arxiv}. A by-product of the change $\phi\to\phi+2\pi$ is that the sign of the relative geometric phase also changes, see Eq.\ \eqref{eq:GeomPhase}. We therefore define primed charge-transfer operators $U'$ which are equal to the original operators introduced in Eq. \eqref{eq:fulloperator}, replacing $\phi$ by $\phi+2\pi$, which leads to a sign flip of $\theta^G$ and $\theta^D$ with respect to $U$.
Including the flux echo as described in sequence A and B gives the following,

\noindent Sequence A':
\begin{align}\nonumber
U'\, U\, \gamma_2\ket{0}_\text{M12}\ket{\psi}_\text{M34}=i&\cos\theta^G\ket{1}_\text{M12}\ket{\psi}_\text{M34}\\
    +&\sin\theta^G\ket{0}_\text{M12}\ket{\psi'}_\text{M34}.\label{eq:seqAprime}
\end{align}
Sequence B':
\begin{align}
    U'\, \gamma_2\, U\ket{0}_\text{M12}\ket{\psi}_\text{M34}=
    i&\ket{1}_\text{M12}\ket{\psi}_\text{M34}.\label{eq:seqBprime}
\end{align}
Because of the $4\pi$-periodicity of the spectrum, we can design a flux echo, equivalent to flipping the system parity. It increases the regime with maximal visibility due to the cancellation of the dynamical phase. Also, the outcome becomes insensitive to the operation time scale.
In sequences A' and B', the final state is $\ket{0}_\text{M12}$ with probability $\sin^2\theta^G$ and 0 respectively. Maximal visibility thus occurs for $\theta^G=\pi/2$. 

To make a measure of the discernibility of the outcome of the two sequences, we introduce the sequence visibility $\Lambda$. We define $\Lambda$ as the difference in probability of measuring the state $
\ket{0}_\text{M12}$ after the two sequences where unit visibility corresponds to the ideal situation. Thus, the sequence visibility for sequences A and B is 
\begin{align}\label{eq:LambdaBAtheory}
    \Lambda&=\sin^2(\theta^G+\theta^D)-\sin^2(\theta^D).
\end{align}
For sequences A' and B' the visibility would simply be
\begin{equation}
    \Lambda'=\sin^2(\theta^G),
\end{equation}
due to the cancellation of dynamical phase. The sequence visibility quantifies the degree to which the orders of operations can be distinguished to show the MBS non-Abelian properties.

In a realistic experiment, tuning the additional SC phase contribution for the flux echo is presumably simpler than tuning $\phi$ to the degeneracy point, integer $\phi/(2\pi)$. 
However, inaccuracies and phase fluctuations can play a role, leading to a nonzero dynamical phase. An additional complication is that the relative dynamical phase is dependent on the exact dynamics of the adiabatic transport. In the next section, we approach the problem of minimizing the dynamical phase contribution using APT to study fast adiabatic processes.

\subsection{Deriving adiabatic perturbation theory}\label{sec:deriveAPT}

The adiabatic theorem predicts that a system initialized in an eigenstate $\ket{n(t=0)}$ of the initial Hamiltonian $H(t=0)$ will follow the instantaneous eigenstate $\ket{n(t)}$ of the slowly varying time-dependent Hamiltonian $H(t)$.
The instantaneous eigenstates fulfill the instantaneous Schrödinger equation,
\begin{equation}
    H(t)\ket{n(t)}=E_n(t)\ket{n(t)}.
\end{equation}
Typically, the adiabatic approximation is valid for
\begin{equation}\label{eq:usualcondition}
    \frac{\left|\braket{m(t)}{\frac{\dd n(t)}{\dd t}}\right|}{|E_m(t)-E_n(t)|}=\frac{\left|\bra{m(t)}\frac{\dd H(t)}{\dd t}\ket{n(t)}\right|}{(E_m(t)-E_n(t))^2}\ll 1, \qquad n\neq m.
\end{equation}
However, this is not always a sufficient condition to ensure adiabaticity \cite{Marzlin2004}. Adiabatic perturbation theory (APT) \cite{Rigolin2008} attempts to determine the validity of the adiabatic approximation, describing nonadiabatic corrections. APT has previously been used in a variety of situations, including quench dynamics through a quantum critical point \cite{DeGrandi2010}, quasi-adiabatic Monte Carlo algorithm \cite{Liu2013APT}, as well as corrections to non-Abelian processes involving Majorana exchange \cite{Cheng2011}. Additionally, APT has also inspired Floquet adiabatic perturbation theory \cite{Weinberg2017, Rodriguez-Vega2021, Deng2016APT}. 

APT is based on a perturbative expansion in the small parameter $1/T$ where $T$ is the relevant time scale of the system \cite{Rigolin2008}. In our case, $T$ is the time of a single charge-transfer operation. The APT expansion parameter $1/T$ is not dimensionless as required by perturbation theories and should be compared to a relevant energy scale. In our system, we have two energy scales $\varepsilon_0$ and $w$ whose ratio $x_0=\varepsilon_0/(2w)$ we take to be large. It is therefore not obvious how to a priori choose the proper dimensionless expansion parameter.

In our study of APT, we simultaneously address this issue and find fast adiabatic energy sweeps of the dot energy to perform efficient charge-transfer operations. While our results are specific to the charge-transfer processes, the framework we use is completely general and may be applied to any nondegenerate quantum system. Further work can presumably extend the framework to degenerate systems as well \cite{Rigolin2010}. We begin our treatment by giving a brief overview of APT as presented in Ref.\ \cite{Rigolin2008}. Then, we apply it to the charge-transfer process, addressing the issues due to the dimensionful expansion parameter $1/T$, and studying fast adiabatic paths. 

For a nondegenerate $N$-level quantum system, APT is based on the following ansatz for the time-evolved state \cite{Rigolin2008}:
\begin{equation}\label{eq:APTansatz}
    \ket{\Psi(s)}=\sum_p^\infty \frac{1}{T^p} \sum_{n,m=0}^{N-1}e^{-iT\omega_m(s)}e^{i\xi_m(s)}b_{nm}^{(p)}(s)\ket{n(s)},
\end{equation}
which is given in terms of the dimensionless time $s=t/T$. The quantities $\omega_m(s)$ and $\xi_m(s)$ are the dynamical and geometric phases of the instantaneous state $\ket{m(s)}$,
\begin{align}
    \omega_m(s)&=\int_0^{s} E_m(s)\, \dd s',\\
    \xi_m(s) &= i \int_0^{s} \braket{m(s')}{\frac{\dd m(s')}{\dd s'}} \dd s'.
\end{align}
The expansion in Eq.\ \eqref{eq:APTansatz} introduces complex, time-dependent coefficients $b_{nm}^{(p)}(s)$ to be determined. Due to the dimensionful expansion parameter $1/T$, the coefficients also carry dimensions such that $b_{nm}^{(p)}(s)/T^p$ is dimensionless. The ansatz in Eq.\ \eqref{eq:APTansatz} recasts the problem of solving the time-dependent Schrödinger equation,
\begin{equation}\label{eq:timeSE}
    \frac{i}{T}\frac{\dd }{\dd s}\ket{\Psi(s)}=H(s)\ket{\Psi(s)},
\end{equation}
into computing the coefficients $b_{nm}^{(p)}(s)$ from linear, recursive equations. The initial conditions for the coefficients are determined by the initial state. In the expansion, the zeroth-order terms correspond to the adiabatic approximation at all times,
\begin{equation}
    b_{nm}^{(0)}(s)=0, \qquad n\neq m.
\end{equation}
It further implies that the initial state is described by the adiabatic approximation, giving the initial constraint on the $p\geq1$ order coefficients,
\begin{equation}
    \sum_m b_{nm}^{(p)}(0)=0, \qquad p\geq1.
\end{equation}
By inserting the ansatz in Eq.\ \eqref{eq:APTansatz} into the time-dependent Schrödinger equation \eqref{eq:timeSE} and taking the inner product with $\bra{m(s)}$ we get,
\begin{align}\nonumber
    i\Delta_{nm}(s)b_{nm}^{(p+1)}(s)+\dot b_{nm}^{(p)}(s)+W_{nm}(s)b_{nm}^{(p)}(s)\\
    +\sum_{k\neq n} M_{nk}(s)b_{km}^{(p)}(s)&=0.\label{eq:rigolinmain}
\end{align}
The following quantities have been defined,
\begin{align} \label{eq:delta}
    \Delta_{nm}(s)&=E_n(s)-E_m(s),\\
    M_{nm}(s)&=\braket{n(s)}{\dot m(s)}=\frac{\bra{n(s)}\dot H(s)\ket{m(s)}}{\Delta_{mn}(s)},\\
    W_{nm}(s)&=M_{nn}(s)-M_{mm}(s), \label{eq:W}
\end{align}
where the dot denotes time differentiation, $\dd/\dd s$. Eq. \eqref{eq:rigolinmain} is the main result of Ref.\ \cite{Rigolin2008} from which the coefficients of order $p+1$ can be recursively computed from the $p$-order coefficients. 

For illustration purposes, we compute the first-order correction in a two-level system initialized in the ground state. Using the initial condition $b_{00}^{(0)}(0)=1$, the first-order coefficients are
\begin{align} \label{eq:firstordercoefficient01}
    b_{01}^{(1)}(s)&=0,\\
    b_{10}^{(1)}(s)&=\frac{iM_{10}(s)}{\Delta_{10}(s)},\\
    b_{00}^{(1)}(s)&=i\int_0^s \frac{|M_{10}(s')|^2}{\Delta_{10}(s')} \, \dd s',\\
    b_{11}^{(1)}(s)&=-\frac{iM_{10}(0)}{\Delta_{10}(0)} \label{eq:firstordercoefficient11}
\end{align}
These first-order coefficients will be the starting point of the next section where we apply APT to the charge-transfer process. We find the optimal adiabatic path and investigate what conditions must be satisfied to be consistent with the adiabatic approximation.

\subsection{Applying adiabatic perturbation theory}\label{sec:applyAPT}

\begin{figure*}
    \centering
    \begin{subfigure}
        \centering
        \includegraphics[height=4.6cm]{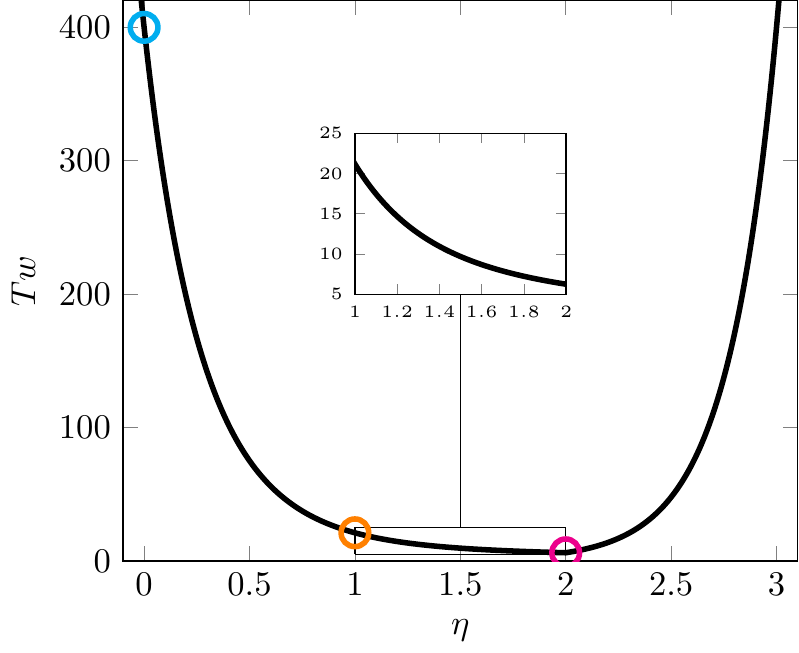}
    \end{subfigure}
    \hfill
    \begin{subfigure}
        \centering
        \includegraphics[height=4.6cm]{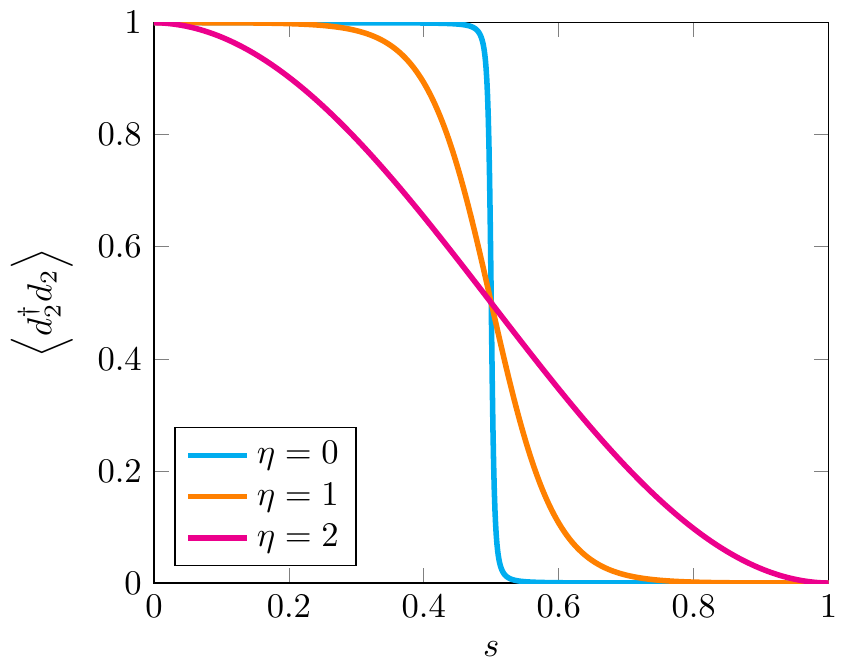}
    \end{subfigure}
    \hfill
    \begin{subfigure}
        \centering
        \includegraphics[height=4.6cm]{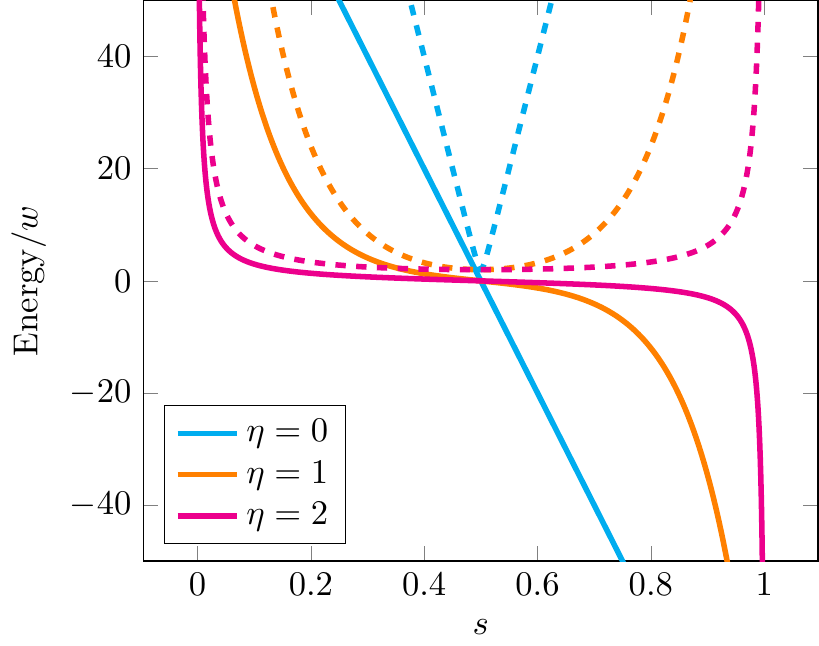}
    \end{subfigure}
    \caption{
    Characteristics of a single charge-transfer process at the degeneracy point (integer $\phi/(2\pi)$) for different values of $\eta$ and $x_0=100$. \textbf{Left:} Operation time scale $T$ (relative to $w$) as a function of $\eta$ for fixed dimensionless expansion parameter $\Sigma_\eta/(Tw)=0.5$, see Eqs.\ \eqref{eq:sigmaeta} and \eqref{eq:Omegaapp}. The panel shows an optimal region for $1<\eta\leq 2$ with an optimal point $\eta=2$, where the adiabatic time scale is the minimal. Colored markers at $\eta=0,1,2$ are reference for the middle and right panels. \textbf{Middle:} Dot occupation $\expval{d_2^\dagger d_2}=\partial E_-/\partial \varepsilon_2$ as a function of dimensionless time $s=t/T$. For the optimal path ($\eta=2$), charge is smoothly transferred during the entire process. For the linear sweep ($\eta=0$), charge is transferred only near the half-way point of the process ($s\approx1/2$), necessitating a longer operation time to ensure adiabatic charge-transfer. \textbf{Right:} Energy sweeps $\varepsilon_2(s)$ (solid lines) and excitation energies $\Delta_{10}(s)$ (dashed lines). For the optimal path ($\eta=2$), most of the operation time is spend where the gap is smallest to avoid nonadiabatic errors. For the linear sweep ($\eta=0$), most of the operation time is spend where the gap is large, leading to a large time scale of the process.
    }
    \label{fig:Sigma}
\end{figure*}

We continue our study by applying APT to the two-level system given in Eq. \eqref{eq:HamMatrix}, which describes two MBSs coupled to a quantum dot.
We use Eqs. (\ref{eq:delta}-\ref{eq:W}) to compute the relevant quantities in the expansion $\Delta_{10}(s)=-\Delta_{01}(s), M_{10}(s)=-(M_{01}(s))^*, W_{10}(s)=-W_{01}(s)$. At the degeneracy point $\sin(2\theta)\sin(\phi/2)\ll1$ we find,
\begin{align}\label{eq:deltaspecific}
    \Delta_{10}(s)&=2w\sqrt{x(s)^2+1},\\ \label{eq:Mspecific}
    M_{10}(s)&=\frac{\dot x(s)}{2(x(s)^2+1)},\\
    W_{10}(s)&=0.
\end{align}
We have expressed the above quantities in terms of the dimensionless level energy $x(s)=\varepsilon_2(s)/(2w)$. Notice that $M_{10}(s)$ is dimensionless and $\Delta_{10}(s)$ has dimension of energy.

To find fast adiabatic paths, we minimize the first-order coefficient $b_{00}^{(1)}(s)$, describing the leading correction to the adiabatic evolution. That is, we minimize the integral
\begin{align}\label{eq:Irho}
    I(s)&=\frac{1}{T}\int_0^s \frac{|M_{10}(s')|^2}{\Delta_{10}(s')}\, \dd s',\\ \label{eq:I}
    &=\frac{1}{8Tw}\int_0^s \frac{\dot x(s')^2}{(x(s')^2+1)^{5/2}}\, \dd s'.
\end{align}
We choose to minimize this coefficient as it describes the nonadiabatic corrections accumulated during the operation. We could also have considered $b_{10}^{(1)}(s)$ or $b_{11}^{(1)}(s)$ which depend on the instantaneous configuration. Before APT, a condition corresponding to $b_{10}^{(1)}(s)$ and Eq. \eqref{eq:usualcondition} was heuristically chosen to find the so-called local adiabatic evolution \cite{Roland2001,Schaller2006,Rezakhani2009}. By minimizing Eq.\ \eqref{eq:I}, we find the optimal adiabatic energy sweep $x_\text{opt}(s)$. Later, we check whether the found adiabatic path is consistent with APT, i.e. the magnitude of the coefficients decrease with the order $p$ and do not grow with $x_0\gg1$.

The integral in Eq.\ \eqref{eq:I} is straightforward to minimize by standard methods. Using the Beltrami identity, we find that the optimal path fulfills
\begin{equation}\label{eq:dotansatz}
    \dot x_\text{opt}(s)=\pm\Omega_\eta\left[x_\text{opt}(s)^2+1\right]^{\eta/2}\propto [\Delta_{10}(s)]^\eta,
\end{equation}
where the $\pm$ sign in front corresponds to emptying or filling the dot and $\Omega_\eta>0$ is a constant dependent on the initial conditions. The minimization of Eq.\ \eqref{eq:I} leads to $\eta=5/2$ as the ideal adiabatic path. The further analysis below, however, shows that $\eta=5/2$ is not optimal as higher-order coefficients are significant for this $\eta$ value. 
In the following of the section, we find the optimal $\eta$ value in Eq. \eqref{eq:dotansatz} consistent with APT constraints. Eq.\ \eqref{eq:dotansatz} is the simplest parametrization which can be physically motivated: the speed of the dot level sweep is proportional to the energy gap between the ground and excited state raised to a power. The energy sweep and the energy gap for $\eta=0,1,2$ is displayed in the right panel of Fig.\ \ref{fig:Sigma}. The case $\eta=0$ corresponds to a linear energy sweep of the quantum dot, independent from the gap to the excited state. $\eta>0$ describes an increasing energy speed of the dot with the gap between the ground and the excited states.
APT also allows to describe more general ansatzes than the one in Eq. \eqref{eq:dotansatz}.

The solution to Eq.\ \eqref{eq:dotansatz} can be given in terms of the Gaussian hypergeometric function $_2 F_1(a,b;c;z)$, see Appendix \ref{sec:hypergeometric}. This enables us to compute the scaling of $\Omega_\eta$ to leading order in $ 1/x_0$ for $x_0\gg1$,
\begin{equation}\label{eq:omegaeta}
	\Omega_\eta\approx\begin{cases}
		\frac{\sqrt{\pi}\Gamma\left(\frac{\eta-1}{2}\right)}{\Gamma\left(\frac{\eta}{2}\right)} & \text{for } \eta>1,\\
		2\sinh[-1](x_0) & \text{for } \eta=1,\\
		\frac{2}{1-\eta}x_0^{1-\eta} & \text{for } \eta<1.
	\end{cases}
\end{equation}
We provide the complete analytic expressions in Appendix \ref{sec:hypergeometric}.
Importantly, $\Omega_\eta$ scales with $x_0$ for $\eta\leq 1$. It can be problematic for APT when evaluating Eq.\ \eqref{eq:I} at $s=1$ in the limit $x_0\gg 1$. Using $x_\text{opt}(s)$ from Eq.\ \eqref{eq:dotansatz},
\begin{equation}\label{eq:Ievaluated}
	I(1)=\frac{\Omega_\eta}{8Tw}\frac{\sqrt{\pi}\Gamma(\frac{4-\eta}{2})}{\Gamma(\frac{5-\eta}{2})},\qquad \text{for } \eta<4.
\end{equation}
A necessary (but insufficient) condition for APT to hold is $I(1)\ll1$, or equivalently, $Tw\gg \Omega_\eta$. It means that for $\eta< 1$, $Tw\gg x_0^{1-\eta}$, which thus requires very slow processes to achieve adiabaticity. For $\eta=1$, $\Omega_\eta$ scales logarithmically with $x_0$. For $\eta>4$, Eq.\ \eqref{eq:I} scales as $x_0^{\eta-4}$. This analysis tells us that for $1<\eta<4$, we need $Tw\gg 1$ to satisfy $I(1)\ll1$. Outside this range, $T$ scales with $x_0$, meaning that the total time for to complete the operation is sensitive to the large energy $\varepsilon_0$. We may also check that the other first-order corrections are small,
\begin{equation}\label{eq:cond2firstorder}
    \frac{M_{10}(s)}{T\Delta_{10}(s)}=\frac{\Omega_\eta}{4Tw}(x_\text{opt}(s)^2+1)^{\frac{\eta-3}{2}}\ll1,
\end{equation}
which decreases with $x_0$ for $\eta<3$ and grows as $x_0^{\eta-3}$ for $\eta>3$, introducing a further restriction to APT validity: $\eta<3$. 
In summary, this preliminary analysis suggests that the first-order corrections are small for $Tw\gg 1$ when $1<\eta<3$. If $\eta$ is chosen outside this range, $T$ grows with $x_0\gg1$. In the following, we show that it is insufficient to demand that the first-order corrections are small for APT to be applicable. This was not mentioned in Ref.\ \cite{Rigolin2008}, but the sufficient conditions are nevertheless contained in APT. Like in the above analysis, we find that $Tw\gg1$ is sufficient but only in the range $1<\eta\leq2$. Outside of this range, large $x_0$ values can make higher-order contributions more significant than the lowest ones in the expansion in Eq. \eqref{eq:APTansatz}. As exemplified in Eq.\ (\ref{eq:Ievaluated}, \ref{eq:cond2firstorder}), this is due to the $w$ and $\varepsilon_0$ dependence of the dimensionful coefficients resulting from the dimensionful expansion coefficient. To resolve this, we express the coefficients in \eqref{eq:APTansatz} of order $p+1$ in terms of $p$-order coefficients,
\begin{align}\nonumber
    b_{nm}^{(p+1)}(s) &= \frac{i}{\Delta_{nm}(s)}\frac{\dd}{\dd s} b_{nm}^{(p)}(s)\qquad (n\neq m)\\
    &+\sum_{k\neq n} \frac{iM_{nk}(s)}{\Delta_{nm}(s)}b_{km}^{(p)}(s),\label{eq:nneqm}
\end{align}
\begin{align}
     \nonumber
    b_{nn}^{(p+1)}(s) &= \sum_{k\neq n}\int_0^s\frac{i M_{nk}(s')}{\Delta_{nk}(s')}\frac{\dd}{\dd s'} b_{kn}^{(p)}(s')\,\dd s'\qquad (n=m)\\ \nonumber
    &+\sum_{\substack{k\neq n\\
    l\neq k}} \int_0^s\frac{iM_{nk}(s')M_{kl}(s')}{\Delta_{nk}(s')}b_{ln}^{(p)}(s')\, \dd s'\\
    &-\sum_{k\neq n}b_{nk}^{(p+1)}(0).\label{eq:neqm}
\end{align}
We demand that the sum of the magnitude of the coefficients of order $p+1$ should be smaller than the corresponding sum of order $p$,
\begin{equation}\label{eq:sumcond}
    \sum_n \sum_m \frac{|b^{(p+1)}_{nm}(s)|}{T^{p+1}}\ll\sum_n \sum_m \frac{|b^{(p)}_{nm}(s)|}{T^{p}}.
\end{equation}
In Appendix \ref{sec:deriveadicond}, we insert Eqs.\ (\ref{eq:nneqm}, \ref{eq:neqm}) into Eq.\ \eqref{eq:sumcond} and get the following adiabatic convergence criteria,
\begin{align}\label{eq:cond1}
     \frac{\Omega_\eta (x_\text{opt}(s)^2+1)^\frac{\eta-1}{2}}{T\Delta_{10}(s)}&\ll1,\\ \label{eq:cond2}
     \frac{|M_{10}(s)|}{T\Delta_{10}(s)}&\ll1,
     \\ \label{eq:cond3}
     \int_0^s\Omega_\eta (x_\text{opt}(s')^2+1)^\frac{\eta-1}{2}\frac{|M_{10}(s')|}{T\Delta_{10}(s')}\, \dd s'&\ll1,\\ \label{eq:cond4}
    \int_0^s \frac{|M_{10}(s)|^2}{T\Delta_{10}(s)}\,\dd s'&\ll1. 
\end{align}
Notice that Eq.\ \eqref{eq:cond2} is identical to the usual adiabatic condition in Eq.\ \eqref{eq:usualcondition}. Furthermore, Eqs.\ \eqref{eq:cond2} and \eqref{eq:cond4} correspond to the conditions found in the first-order coefficients in Eqs.\ \eqref{eq:Ievaluated} and \eqref{eq:cond2firstorder}. Our extended analysis in Appendix \ref{sec:deriveadicond} have thus provided two additional conditions to satisfy adiabaticity, Eqs.\ \eqref{eq:cond1}, \eqref{eq:cond3}. The additional conditions come from terms in Eqs.\ (\ref{eq:nneqm}-\ref{eq:neqm}) which do not appear when computing the first-order coefficients but become relevant in higher-order ones.

In the regime $|x_\text{opt}(s)|\sim 1$, the conditions (\ref{eq:cond1}-\ref{eq:cond4}) result in $\Omega_\eta/(Tw)\ll1$, which gives the lower bound $\eta> 1$ as discussed above. For large $|x_\text{opt}(s)|$, the convergence of the integral in Eq.\ \eqref{eq:cond3} gives the upper bound $\eta<3$ which was the same as in the conditions \eqref{eq:cond2firstorder} and \eqref{eq:cond2}. Importantly, the first condition \eqref{eq:cond1} gives a further restriction for large $|x_\text{opt}(s)|$,
\begin{equation}\label{eq:cond1scaling}
    \frac{\Omega_\eta}{Tw}x_0^{\eta-2}\ll  1
\end{equation}
This is the final restriction on $\eta$ and gets us the bound for optimal operation time $Tw\gg1$,
\begin{equation}\label{eq:etabound}
    1<\eta \leq 2.
\end{equation}
We note that both the linear energy sweep ($\eta=0$) and the best adiabatic path ($\eta=5/2$) predicted by the first-order correction in Eq.\ \eqref{eq:I} lie outside the optimal range.

To make an unified statement about the proper dimensionless expansion parameter, we define a quantity closely related to $\Omega_\eta$, including the scaling for $\eta>2$,
\begin{equation}\label{eq:sigmaeta}
    \Sigma_\eta=\begin{cases}
        \Omega_\eta x_0^{\eta-2} & \text{for } \eta>2,\\
		\Omega_\eta & \text{for } \eta\leq2.
		\end{cases}
\end{equation}
We thus propose $\Sigma_\eta/(Tw)$ as the proper dimensionless expansion parameter, fulfilling $\Sigma_\eta/(Tw)\ll1$ for APT to hold. This expansion parameter depends in a nontrivial way on $w$ and $\varepsilon_0$ and the chosen path parametrized by $\eta$.

\begin{figure*}
    \centering
    \begin{subfigure}
        \centering
        \includegraphics[height=6cm]{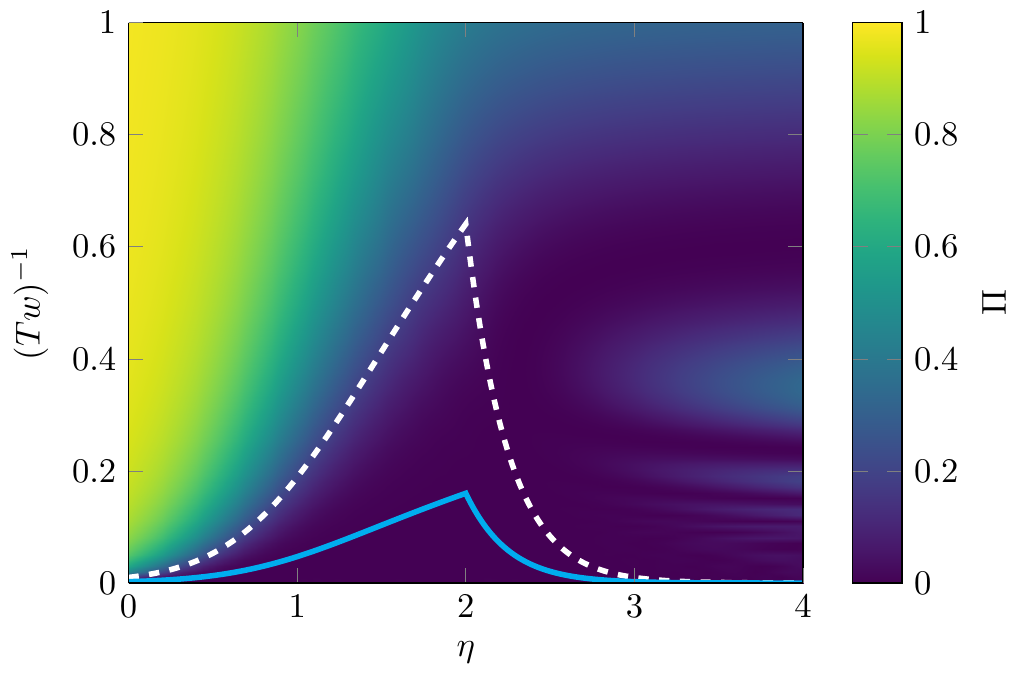}
    \end{subfigure}
    \hfill
    \begin{subfigure}
        \centering
        \includegraphics[height=6cm]{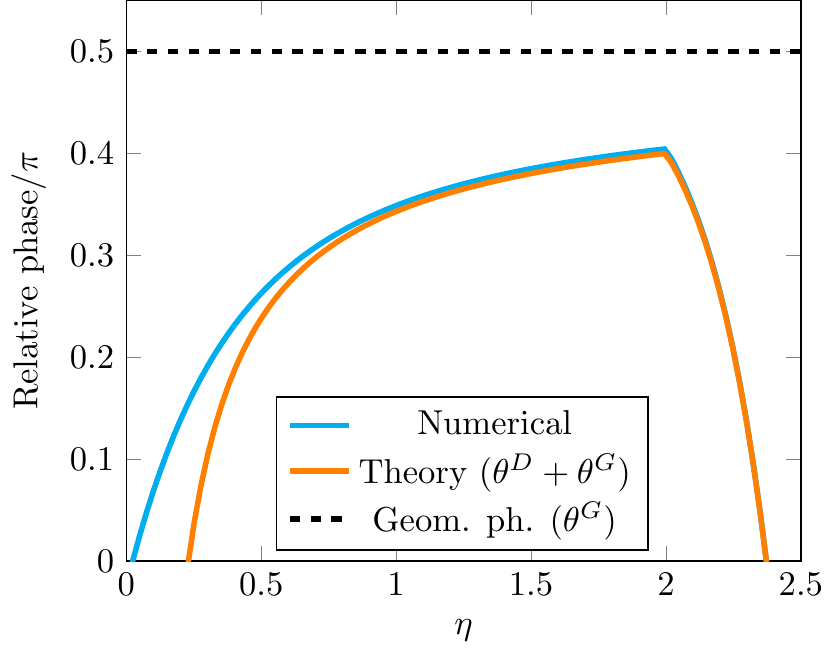}
    \end{subfigure}
    \caption{Numerical results for a charge-transfer process with $\theta=\pi/4$ and $x_0=100$.
    \textbf{Left:} Color map of the transition probability after a single charge-transfer operation at the degeneracy point ($\phi=0$) as a function of $\eta$ and the inverse time $(Tw)^{-1}$. The two lines represent the prediction from APT for $\Sigma_\eta/(Tw)=2$ (dashed white) and $\Sigma_\eta/(Tw)=0.5$ (solid cyan). For $\eta\leq 2$, the dashed white line separates the adiabatic region (dark blue) from the nonadiabatic region (green and yellow). The solid cyan line lies well in the adiabatic region and is used for reference to the right panel.
    \textbf{Right:} Plot of the relative phase for a slight detuning $\phi=0.05\pi$ from the ideal phase ($\phi=0$) following the cut at the solid cyan line in the left panel ($\Sigma_\eta/(Tw)=0.5$).
    We display the numerical result (cyan), theoretical prediction (orange) and the geometric phase (dashed) for reference to the ideal situation.}
    \label{fig:transitionandphase}
\end{figure*}

APT predicts that the fastest adiabatic path is the solution to Eq.\ \eqref{eq:dotansatz} for $\eta=2$, which minimizes the dimensionless expansion parameter $\Sigma_{\eta=2}/(Tw)=\pi/(Tw)$, see left panel of Fig.\ \ref{fig:Sigma}. For $\eta=2$, the solution to Eq.\ \eqref{eq:dotansatz} has a particularly simple expression given by
\begin{equation}\label{eq:optimalpath}
    x_\text{opt}(s)=\pm\tan[\arctan(x_0)\left(2s-1\right)].
\end{equation}
This result realizes the so-called local adiabatic evolution of the system \cite{Roland2001,Schaller2006,Rezakhani2009}.
In Fig.\ \ref{fig:Sigma} the optimal sweep ($\eta=2$) is compared to a linear sweep ($\eta=0$). The ratio $\Sigma_{\eta=0}/\Sigma_{\eta=2}\approx2x_0/\pi$ quantifies how much faster the optimal sweep of $x_\text{opt}(s)$ can be with respect to a linear one. This means that, for the same parameters, the ideal sweep is $\approx 64$ times faster than the linear one for $x_0=100$. The intuition is that the charge is exchanged at a nearly constant rate for $\eta=2$, see middle panel of Fig.\ \ref{fig:Sigma}. However, the system spends most of the time in a region where no charge is transferred for $\eta=0$. Finally, using Eq.\ \eqref{eq:dotansatz}, we compute the relative dynamical phase considered in Sec.\ \ref{sec:charge-transfer-process} to first-order in $\sin(2\theta)\sin(\phi/2)$ and in the limit $x_0\gg1$,
\begin{align}\nonumber
    \theta^D&=-T\int_0^1 (E_-^+(s)-E_-^- (s))\, \dd s ,\\
    &=-\sin(2\theta)\sin(\phi/2)\, \frac{\sqrt{\pi}\, \Gamma(\frac{\eta}{2})}{\Gamma\left(\frac{\eta+1}{2}\right)}\, \frac{Tw}{\Omega_\eta}. \label{eq:DynPhase}
\end{align}
This equation describes a decreasing undesired dynamical phase when $\eta$ increases. This further motivates the choice $\eta=2$ for the charge-transfer process.

We conclude this section by outlining the presented framework for finding fast adiabatic paths while checking adiabatic conditions. The method can be broken down into the following five steps: 
\begin{enumerate}
    \item Write down the first-order corrections using APT, Eqs.\ (\ref{eq:firstordercoefficient01}-\ref{eq:firstordercoefficient11}).
    \item From the first-order coefficients, choose a relevant functional, Eqs.\ \eqref{eq:Irho} and \eqref{eq:I}, and minimize it.
    \item Extend the family of considered paths by parametrizing the minimizing differential equation, Eq.\ \eqref{eq:dotansatz}.
    \item Check the adiabatic conditions, constraining the parameters, Eqs.\ \eqref{eq:omegaeta} and (\ref{eq:nneqm}-\ref{eq:etabound}).
    \item Choose the set of parameters that minimizes the proper dimensionless expansion parameter, Eq.\ \eqref{eq:sigmaeta}. The path obtained through this procedure, Eq.\ \eqref{eq:optimalpath}, is the optimal adiabatic one for the family considered in step 3.
\end{enumerate}

This procedure thus provides an optimal adiabatic path, taking into account nonadiabatic corrections. The framework is general and may be used to find fast adiabatic paths in other systems. Future efforts may also expand the framework to include degenerate quantum systems \cite{Rigolin2010}.

In general, higher time-derivatives of the Hamiltonian at $s=0$ and $s=1$ can lead to additional nonadiabatic contributions not captured by APT. We have not considered these effects as they appear to play a minor role due to the large initial and final energy gaps between the ground and excited states. In the case where these gaps are comparable to other energy scales in the system, the contributions from the higher time-derivatives of the Hamiltonian
can have some influence in the result. In this case, boundary cancellation techniques can be used to reduce such contributions \cite{Passos2020}. Finally, we would like to mention the existence of methods exploiting symmetry to improve the error-scaling \cite{Wiebe2012, Posske2020}. It may further reduce the time scale of the charge-transfer process.

\begin{figure*}
    \centering
    \begin{subfigure}
        \centering
        \includegraphics[height=5.8cm]{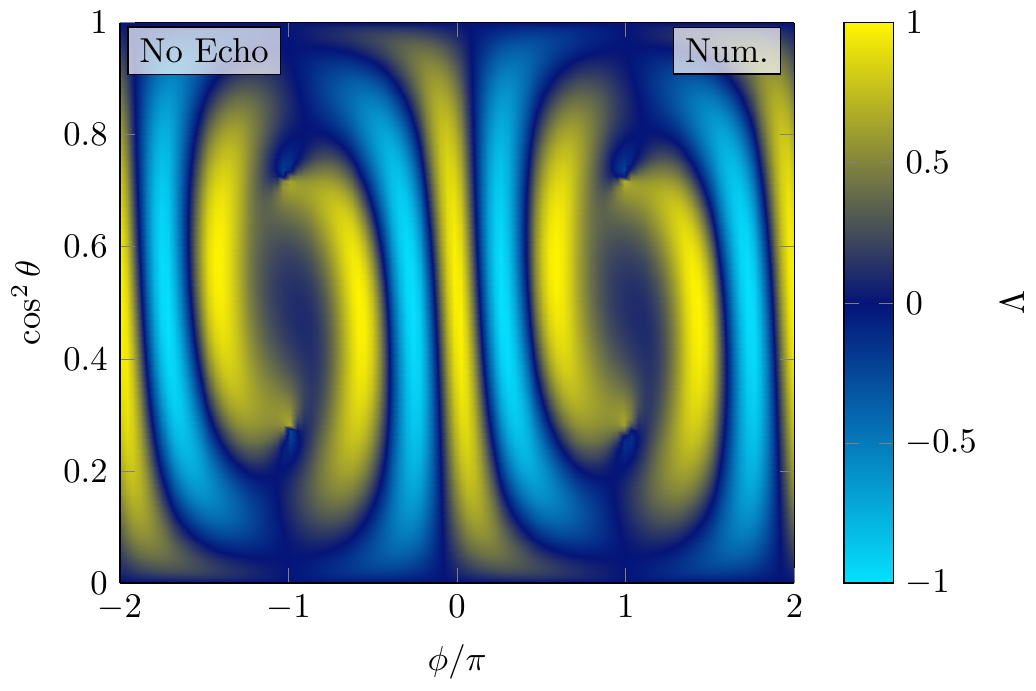}
    \end{subfigure}
    \hfill
    \begin{subfigure}
        \centering
        \includegraphics[height=5.8cm]{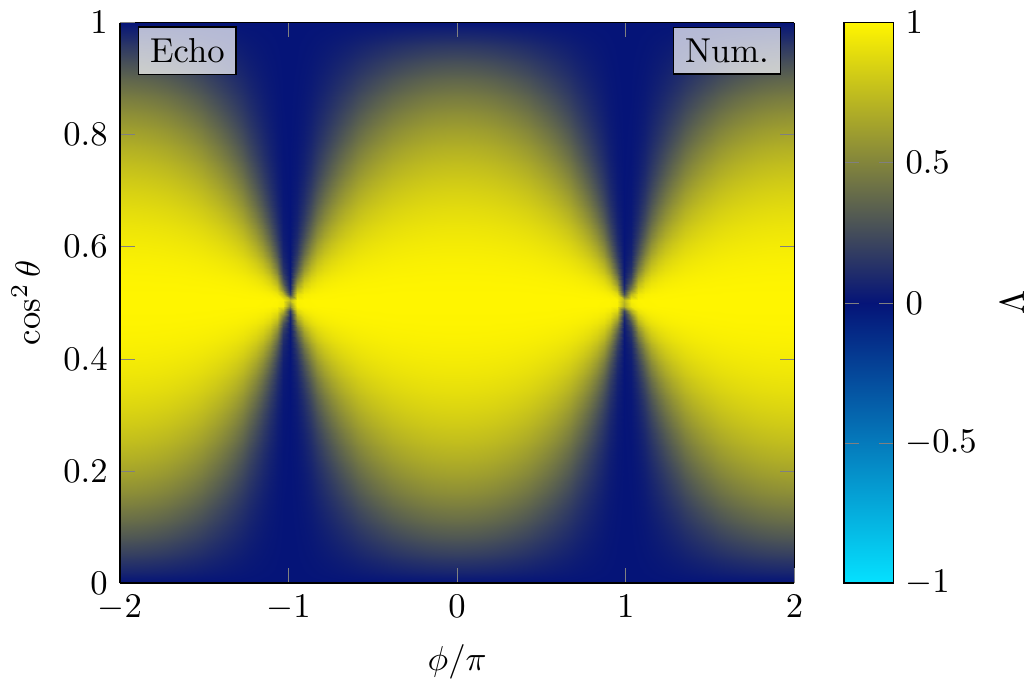}
    \end{subfigure}
    
    \begin{subfigure}
        \centering
        \includegraphics[height=5.8cm]{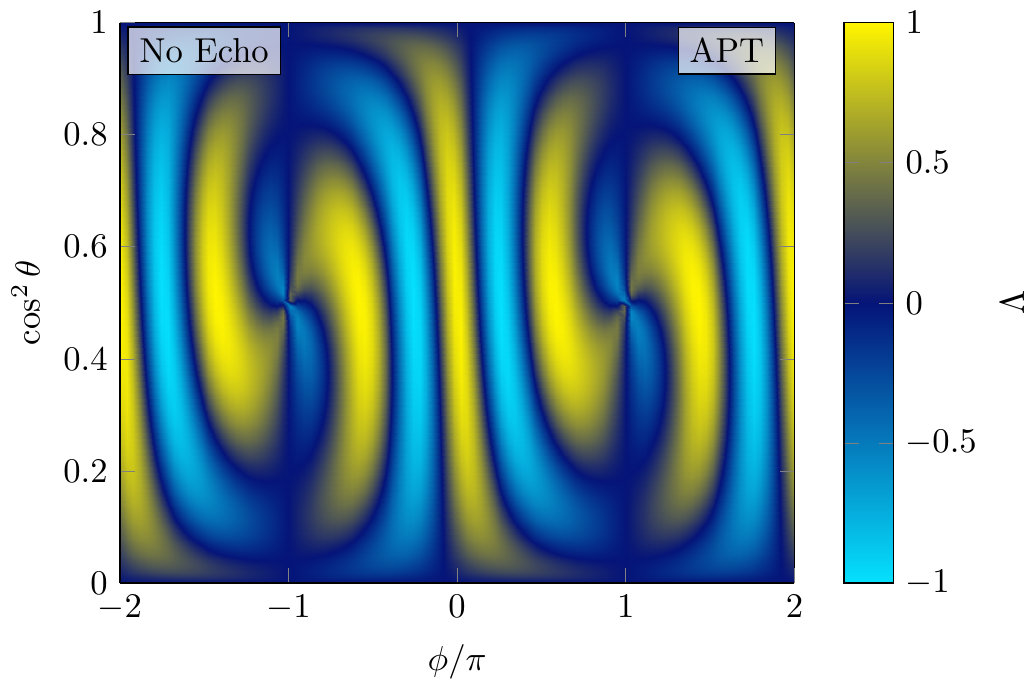}
    \end{subfigure}
    \hfill
    \begin{subfigure}
        \centering
        \includegraphics[height=5.8cm]{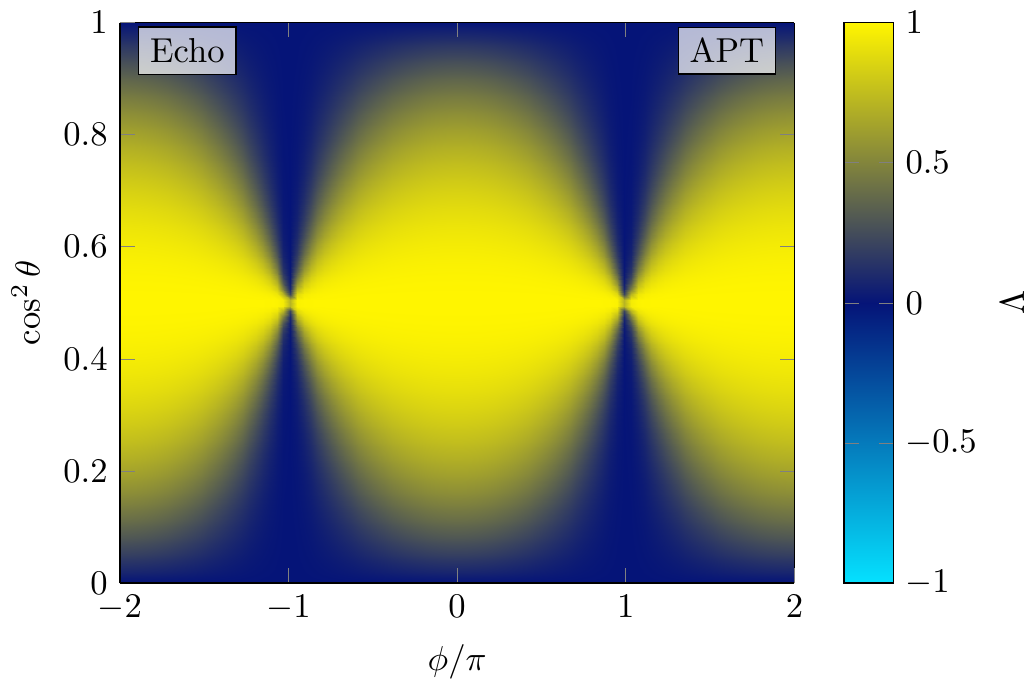}
    \end{subfigure}
    \caption{Sequence visibility, Eq. \eqref{eq:LambdaBAtheory}, as a function of the initial detuning $\phi$ and the coupling strength asymmetry $\cos^2\theta$. We compare numerical simulations of the protocol proposed in Sec. \ref{sec:protocol} (top panels) and APT predictions (bottom panels). We show results with (right panels) and without the flux echo protocol (left panels).}
    \label{fig:mainresult}
\end{figure*}

\section{Numerical results}

In this section, we test the predictions of APT numerically. We show that the dimensionless expansion parameter $\Sigma_\eta/(Tw)$ describes the adiabatic condition. We pick an optimal path based on the APT prediction, which minimizes the operation time scale and the nonadiabatic errors. We simulate numerically the protocol with and without the flux echo. We find that the echo protocol substantially extends the parameter space where MBS non-Abelian properties can be shown using charge-transfer operations.

In the left panel of Fig.\ \ref{fig:transitionandphase}, we display the probability of transitioning to the excited state, $\Pi$, as a function of $\eta$ and the inverse operation time, $(Tw)^{-1}$. We show results after a single charge-transfer operation in the case where the even and odd parity sectors are degenerate. As expected, the transition probability to the excited state decreases when the operation time increases. The white line is a contour of the dimensionless expansion parameter, $\Sigma_\eta/(Tw)=2$. As suggested from APT, the dimensionless expansion parameter separates well the adiabatic (suppressed $\Pi$ region below the line) and the nonadiabatic regimes (larger $\Pi$ region above the line). APT agrees quantitatively with the numerical calculations for $\eta\leq 2$. For $\eta>2$, the contour avoids the regions of nonzero transition probability in the lower right corner. In this region outside of the APT regime, the system behavior is non-monotonic, as shown by the local $\Pi$ maxima as a function of the operation time. In Appendix \ref{sec:deriveadicond}, we further discuss the APT prediction at $\eta>2$.

The solid cyan line, given by $\Sigma_\eta/(Tw)=0.5$, lies in the adiabatic region, where charge-transfer operations can be done with high accuracy.
In the right panel of Fig.\ \ref{fig:transitionandphase}, we show the relative phase between the even and odd ground states after a single charge transfer operation following the cyan line in the left panel for $\phi=0.05\pi$. 
For charge-transfer operations, small deviations from the ideal conditions can lead to a significant relative dynamical phase as illustrated by the difference between the dashed (ideal result) and the solid lines.
The agreement between the numerical result and APT is good, except close to $\eta=0$. This is due to the approximation $x_0\to \infty$ when computing $\theta^D$ in Eq.\ \eqref{eq:DynPhase}. 

Combining the results obtained by the numerically simulated charge-transfer operations, we conclude that $\Sigma_\eta/(Tw)\simeq0.5$ and $\eta=2$ are the best values, as suggested by APT. As for realistic parameters, we assume that the induced superconducting gap is $\Delta_{\text{SC}}=0.1$ meV. To avoid transitioning to the continuum of states, we take $\varepsilon_0=0.5\, \Delta_\text{SC}=50 \ \mu$eV. Using a value of $x_0=\varepsilon_0/(2w)=100$, we get $w=0.25\ \mu$eV and $T\approx 17 $ ns. It is thus possible to perform fast adiabatic charge-transfer operations on the $\sim10$ ns scale. The transition probability for these parameters is $\Pi<10^{-5}$. Using the same parameters, but with a linear sweep ($\eta=0$), the corresponding time scale is approximately $1\ \mu$s with similar transition probability. Previous experiments have shown that parity lifetime in trivial superconducting islands are $\sim1\ \mu$s \cite{Albrecht2017}, illustrating that it might not be possible to perform accurate operations using a linear sweep.

Using the optimal path found, $\Sigma_\eta/(Tw)=0.5$ and $\eta=2$, we simulate the protocols described in Sec.\ \ref{sec:protocol} to demonstrate MBS non-Abelian properties. The results are shown in Fig.\ \ref{fig:mainresult}. Here we make color maps of the sequence visibility $\Lambda$ as a function of $\phi$ and the coupling asymmetry $\cos^2\theta$. As explained around Eq.\ \eqref{eq:LambdaBAtheory}, $\Lambda$ measures how well the sequences in Eqs. (\ref{eq:seqA}-\ref{eq:seqBprime}) can be distinguished by the measured parity of the M12 fermion. It thus quantifies the confidence of demonstrating non-Abelian properties. Here, $\Lambda=\pm1$ means that the parity of M12 fermion can distinguish between the two sets of operations, while the protocol fails for $\Lambda=0$.

In the top left panel of Fig.\ \ref{fig:mainresult}, we display numerical results for the visibility for the protocol without the echo. Note that the optimal parameter values $\theta=\pi/4$ and $\phi=0$ lie at the central yellow sliver with maximal visibility. The narrow width ($\approx 0.1\pi$) of this high-visibility region is due to the contribution of the dynamical phase and illustrates the importance of accurately tuning $\phi$. It appears less important to tune the coupling asymmetry $\theta$. In Appendix \ref{app:Visibilitytime scales}, we display the sequence visibility for different $T$ values to show that the width of the high-visibility regions decreases as $T$ is increased. The top left panel should be compared to the numerical results for the echo protocol displayed in the top right panel. Here the central yellow region is significantly extended due to the cancellation of the dynamical phase, making the experiment rather insensitive to $\phi$. The outcome is also insensitive to $T$, as shown in Appendix \ref{app:Visibilitytime scales}. The echo protocol, however, depends on tuning $\phi\to \phi+\delta\phi$ with $\delta\phi=2\pi$ ideally and is thus robust to drifts in $\phi$. In Appendix \ref{app:RobustEcho}, we offset the parameter $\delta\phi$ and find that the echo protocol is robust up to deviation of $\sim0.2\pi$ in $\delta\phi$. For completeness, we show  the probability to end up in the state $\ket{0}_\text{M12}$ after each sequence in Appendix \ref{app:SignatureSequence}.

In the bottom panels of Fig.\ \ref{fig:mainresult}, we display the visibility obtained from APT, in good agreement with the numerical results shown in the top row panels. However, there is a discrepancy in the region $\sin(2\theta)\sin(\phi/2)\sim 1$. The disagreement between theory and the numerical results is due to the closing of the gap between the ground and the excited states as $w\sqrt{1-\rho\sin(2\theta)\sin(\phi/2)}$, Eq.\ \eqref{eq:eigenenergy}. This results in transitions and large nonadiabatic errors to the phase in that region.

\section{Conclusions and discussions}
In this work, we have proposed a minimal experiment for demonstrating Majorana non-Abelian properties. The experiment requires three Majorana bound states (MBSs), the minimal number to measure non-Abelian signatures. Our proposal is based on charge-transfer operations between a quantum dot and two MBSs. Another quantum dot is used for the initialization and readout. 
We also devise a minimal protocol relying on two sequences of three adiabatic charge-transfer operations. The final result depends on the order of operations due to Majorana non-Abelian properties.

We study the robustness of the protocol as a function of the model parameters, taking into account nonadiabatic effects. To this end, we develop a framework based on adiabatic perturbation theory (APT) for finding fast adiabatic paths in nondegenerate quantum systems. 
This framework describes the optimal adiabatic energy sweep for the charge-transfer operation. We find that the experiment is sensitive to the SC phase difference, $\phi$. Small deviations, $\sim0.05\pi$ from the degeneracy point ($\phi=0$) lead to a substantial dynamical phase that can dominate over the non-Abelian signal.
To solve this issue, we propose a flux echo protocol that significantly reduces the sensitivity on $\phi$. The flux echo relies on increasing the superconducting phase difference by $2\pi$ between subsequent operations, exploiting the $4\pi$-periodicity of the topological state. The tolerance on the additional phase is $\sim0.2\pi$, while the outcome of the protocol is insensitive to the operation time and robust to drifts in $\phi$.

Since our proposal relies on parameter space operations rather than real space braiding, it is relevant to discuss the uniqueness of the MBS signature in the proposed experiment. A system hosting trivial subgap states may also acquire geometric and dynamical phases during charge-transfer operations. As a result, charge-transfer operations might not commute, leading to potentially large $\Lambda$ values for some parameters. However, the flux echo, exploiting MBSs $4\pi$-periodicity, leads to a robust non-Abelian signal over a wide range of parameters. This is in contrast to trivial bound states, which are $2\pi$-periodic, where large $\Lambda$ values only appear at fine-tuned situations due to the dynamical phase.
Other than trivial states, the experiment might also suffer from various sources of error that can lead to a reduction of the non-Abelian signal. First, fluctuations in the superconducting phase difference will introduce a random phase. However, the flux echo protocol reduces their effect if the operations are faster than the timescale of phase fluctuation. Second, the coupling between MBSs will split the ground state degeneracy introducing a constraint on the upper limit for the charge-transfer operations. However, as shown in Ref.\ \cite{Souto2020}, this effect is likely not a limiting factor. Additionally, quasiparticle poisoning is detrimental to the experiment and its timescale should therefore be longer than that of the experiment. Finally, non-zero temperature and electric fluctuations in the gates will reduce the non-Abelian signal. In these cases, the tunnel coupling strength should be larger than the temperature and electric variations. Also, the optimal path found, minimizing the operation timescale reduces their impact.

\section{Acknowledgments}
This research was supported by the Danish National Research Foundation, the Danish Council for Independent Research $|$ Natural Sciences. This project has received funding from the European Research Council (ERC) under the European Union’s Horizon 2020 research and innovation programme under Grant Agreement No. 856526. We acknowledge support from the Deutsche Forschungsgemeinschaft (DFG) – project grant 277101999 – within the CRC network TR 183 (subproject C03). R.S.S. acknowledges funding from QuantERA project 2D hybrid materials as a platform for topological quantum computing” and from NanoLund.

\appendix

\section{The geometric phase}\label{sec:appendix_1l}
There is a technical subtlety when computing the relative geometric phase in Eq.\ \eqref{eq:GeomPhase}: a single charge-transfer process does not constitute a loop in parameter space. It makes difficult to determine the acquired geometrical phase. We instead compare the geometric phases collected by the even and odd ground states during a charge-transfer process. However, the even and odd parity ground states live in different Hilbert spaces. Since there is a clear one-to-one mapping between these two spaces, we treat the ground state vectors as living in the same Hilbert space. 

The gauge choice in Eq.\ \eqref{eq:eigenstates} is such that for each parity, there is no mathematical contribution to the geometric phase when changing $\varepsilon_2: \varepsilon_0\to-\varepsilon_0$ in time $T$,
\begin{equation}
    i\int_{0}^{T} \dd t\ (\psi_-^\rho)^\dagger \frac{\dd \psi_-^\rho}{\dd t}=0,
\end{equation}
This is easy to see as the ground states have the form $(\psi_-^\rho)^\dagger=\left(e^{i\xi}\cos(\lambda(t)), \sin(\lambda(t))\right)$.
The gauge choice in Eq.\ \eqref{eq:eigenstates}, however, is different for the two parity sectors and this gives a relative geometric phase between the even and odd parity ground states. To compute this relative geometric phase contribution, we evaluate the phase difference between the ground states using $\arctan\left[\frac{\Im[(\psi_-^+)^\dagger\cdot\, \psi_-^-]}{\Re[(\psi_-^+)^\dagger\cdot\, \psi_-^-]}\right]$ and compare the results at initial and final values of the level energy. This calculation leads to the result in Eq.\ \eqref{eq:GeomPhase}.

The relative geometric phase can also be understood as a proper loop in parameter space by noticing that the even and odd parity Hamiltonian and eigenvectors can be transformed into each other by $\theta\to-\theta$. We can thus compute the relative geometric phase by considering the loop $\varepsilon_0\to-\varepsilon_0$, $\theta\to-\theta$, $-\varepsilon_0\to\varepsilon_0$, $-\theta\to\theta$. This can be understood as performing a charge-transfer operation in the even parity state, 
inverting $\theta$ to transform it to the odd parity state. We then perform another operation and invert again the sign of $\theta$ to return to the even subspace.
The geometric phase due to this loop corresponds to the relative geometric phase acquired between the even and odd parity ground states due to a single charge-transfer process. There is no contribution to the geometric phase for large negative level energies as the ground states become $(\psi_-^\rho)^\dagger=(0,-1)$ in this limit. At the other side of the loop, where the level energy has a large positive value, the ground states are $(\psi_-^\rho)^\dagger=(w^\rho/|w^\rho|,0)$. Using the gauge in Eq.\ \eqref{eq:eigenstates} no geometrical phase is acquired by the system when varying $\varepsilon_2$. The relative geometric phase is given by
\begin{equation}\label{eq:BerryPhase}
    \theta^G=i\int_{-\theta}^{\theta}\dd\theta'\ (\psi_-^-)^\dagger \frac{\dd \psi_-^-}{\dd \theta'},
\end{equation}
in the limit of large positive level energies.
This approach provides an alternative picture of how to calculate the geometric phase, but mathematically it is tedious to carry out. Performing the integration in Eq.\ \eqref{eq:BerryPhase} and envoking the identity
\begin{align}\nonumber
    2\arctan(\tan(x)\cos(y))=&\arctan(\frac{\tan{x}}{\cos{y}}-\frac{\tan{y}}{\cos{x}})\\
    +&\arctan(\frac{\tan{x}}{\cos{y}}+\frac{\tan{y}}{\cos{x}}),
\end{align}
we arrive at Eq. \eqref{eq:GeomPhase}.

\section{Solution in terms of the Gaussian hypergeometric function}\label{sec:hypergeometric}

For a symmetric charge-transfer following,
\begin{equation}
    \dot x_\text{opt}(s)=\pm\Omega_\eta\left[x_\text{opt}(s)^2+1\right]^{\eta/2},
\end{equation}
the solution is 
\begin{equation}
    \pm\Omega_\eta (s-1/2)=x_\text{opt}(s)\, _2F_1\left(\frac{1}{2},\frac{\eta}{2};\frac{3}{2};-x_\text{opt}(s)^2\right),
\end{equation}
where the Gaussian hypergeometric function is defined by
\begin{equation}\label{eq:GaussHyper}
	_2F_1(a,b;c;z)=\frac{\Gamma(c)}{\Gamma(a)\Gamma(b)}\sum_n^\infty \frac{\Gamma(a+n)\Gamma(b+n)}{\Gamma(c+n)n!}z^n,\ |z|<1.
\end{equation}
The initial and final conditions determine $\Omega_\eta$,
\begin{equation}\label{eq:Omegaapp}
    \Omega_\eta=2x_0\, _2F_1\left(\frac{1}{2},\frac{\eta}{2};\frac{3}{2};-x_0^2\right).
\end{equation}
To get the approximation for large $x_0$ in Eq.\ \eqref{eq:omegaeta}, we use the transformation rule 
\begin{align}\label{eq:GaussHyperTransform}
	&_2F_1(a,b;c;z)=\\
	&\frac{\Gamma(c)\Gamma(b-a)}{\Gamma(b)\Gamma(c-a)}(-z)^{-a}\, _2F_1(a,a-c+1;a-b+1;1/z)\\
	&+ (a\leftrightarrow b),\qquad \text{for}\ |\arg(-z)|<\pi.
\end{align}

\section{Deriving adiabatic conditions}\label{sec:deriveadicond}

In this section, we derive the adiabatic conditions, Eqs.\ (\ref{eq:cond1}-\ref{eq:cond4}) in the main text, starting from Eqs. (\ref{eq:nneqm}-\ref{eq:sumcond}). We omit in the following the time variable for simplicity.

In Eq.\ \eqref{eq:sumcond}, we split the left hand side term into contributions from $n=m$ and $n\neq m$,
\begin{equation}
    \sum_n \sum_{m} \frac{|b^{(p+1)}_{nm}|}{T^{p+1}}=\sum_n\left( \sum_{m\neq n} \frac{|b^{(p+1)}_{nm}|}{T^{p+1}}+\frac{|b^{(p+1)}_{nn}|}{T^{p+1}}\right).
\end{equation}
The condition \eqref{eq:sumcond} is satisfied if each term is individually smaller than its right hand side,
\begin{align}\label{eq:cond1app}
    \sum_n \sum_{m\neq n} \frac{|b^{(p+1)}_{nm}|}{T^{p+1}}\ll\sum_n \sum_m \frac{|b^{(p)}_{nm}|}{T^{p}},\\ \label{eq:cond2app}
    \sum_n \frac{|b^{(p+1)}_{nn}|}{T^{p+1}}\ll\sum_n \sum_m \frac{|b^{(p)}_{nm}|}{T^{p}}.
\end{align}
We study these two cases separately. We begin with the $n\neq m$ case, substituting Eq.\ \eqref{eq:nneqm} in Eq.\ \eqref{eq:cond1app}
\begin{align}\nonumber
    &\sum_n \sum_{m\neq n} \frac{|b^{(p+1)}_{nm}|}{T^{p+1}}\\
    &=\sum_n \sum_{m\neq n}\left|\frac{i}{T\Delta_{nm}}\frac{\dd}{\dd s} \frac{b_{nm}^{(p)}}{T^p}+\sum_{k\neq n} \frac{iM_{nk}}{T\Delta_{nm}}\frac{b_{km}^{(p)}}{T^p} \right|,\\
    &\leq \sum_n \sum_{m\neq n}\left(\frac{1}{T|\Delta_{nm}|}\left|\frac{\dd}{\dd s} \frac{b_{nm}^{(p)}}{T^p}\right|+\sum_{k\neq n} \frac{|M_{nk}|}{T|\Delta_{nm}|}\frac{|b_{km}^{(p)}|}{T^p}\right). 
\end{align}
Again, the condition \eqref{eq:sumcond} is satisfied if each term fulfills
\begin{align}\label{eq:cond1.1app}
    \sum_n \sum_{m\neq n}\frac{1}{T|\Delta_{nm}|}\left|\frac{\dd}{\dd s} \frac{b_{nm}^{(p)}}{T^p}\right|\ll \sum_n \sum_m \frac{|b^{(p)}_{nm}|}{T^{p}},\\ \label{eq:cond1.2app}
    \sum_n \sum_{m} \left(\sum_{k\neq n,m} \frac{|M_{nk}|}{T|\Delta_{mk}|}\right)\frac{|b_{nm}^{(p)}|}{T^p}\ll\sum_n \sum_m \frac{|b^{(p)}_{nm}|}{T^{p}},
\end{align}
where we have relabelled the sums.
Similarly, by substituting Eq.\ \eqref{eq:neqm} to the left hand side of Eq.\ \eqref{eq:cond2app} and considering each term separately, we get
\begin{align}\label{eq:cond2.1app}
    \sum_n\sum_{m\neq n}\int_0^s \frac{|M_{nm}|}{T|\Delta_{nm}|}\left|\frac{\dd}{\dd s'}b_{nm}^{(p)}\right|\,\dd s'&\ll\sum_n \sum_m \frac{|b^{(p)}_{nm}|}{T^{p}},\\ \label{eq:cond2.2app}
    \sum_n\sum_{m}\int_0^s\left|\sum_{k\neq n,m} \frac{M_{mk}M_{kn}}{T\Delta_{mk}}\right| \frac{|b_{nm}^{(p)}|}{T^p}\,\dd s'&\ll\sum_n \sum_m \frac{|b^{(p)}_{nm}|}{T^{p}},\\ \label{eq:cond2.3app}
    \sum_n\sum_{m\neq n}\frac{|b_{nm}^{(p+1)}(0)|}{T^{p+1}}&\ll\sum_n \sum_m \frac{|b^{(p)}_{nm}|}{T^{p}}.
\end{align}
Note that the last of these conditions is included in Eq.\ \eqref{eq:cond1app}. 

We first focus on Eqs.\ \eqref{eq:cond1.2app} and \eqref{eq:cond2.2app}, which are the simplest inequalities. They are satisfied for
\begin{align} \label{eq:cond1.2.1app}
    \sum_{k\neq n,m} \frac{|M_{nk}|}{T|\Delta_{mk}|}&\ll1, \\ \label{eq:cond2.2.1app}
    \int_0^s\left|\sum_{k\neq n,m} \frac{M_{mk}M_{kn}}{T\Delta_{mk}}\right|\,\dd s'&\ll1.
\end{align}
For a two level system as the one considered in Sec.\ \ref{sec:applyAPT}, Eqs. \eqref{eq:cond1.2.1app} and \eqref{eq:cond2.2.1app} results in the conditions in Eqs.\ \eqref{eq:cond2} and \eqref{eq:cond4}.

\begin{figure*}
    \centering
    \begin{subfigure}
        \centering
        \includegraphics[height=5.8cm]{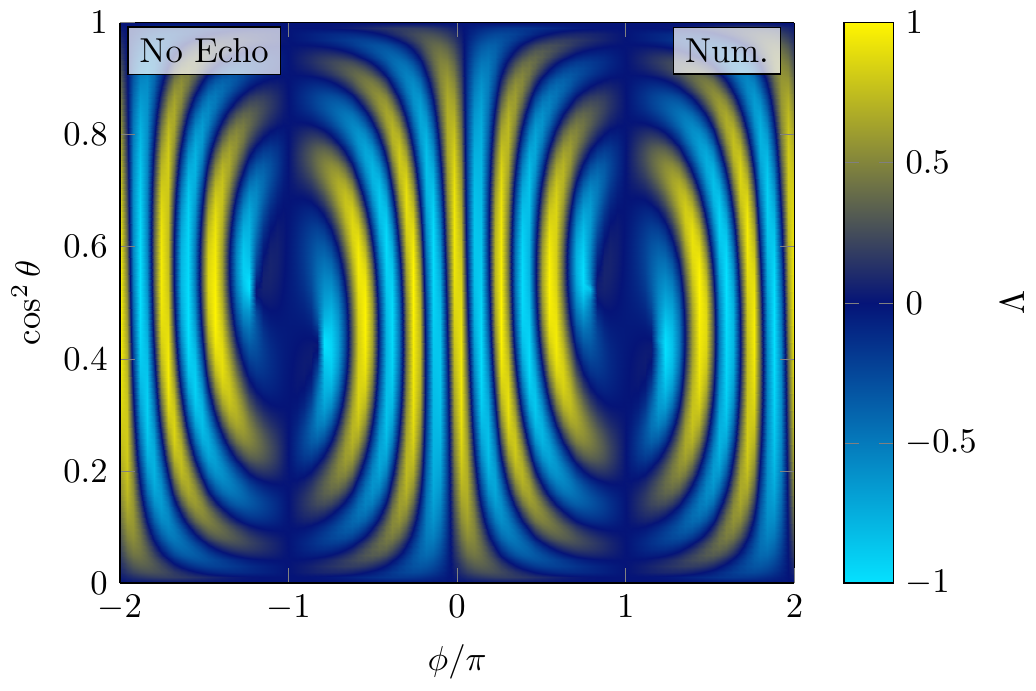}
    \end{subfigure}
    \hfill
    \begin{subfigure}
        \centering
        \includegraphics[height=5.8cm]{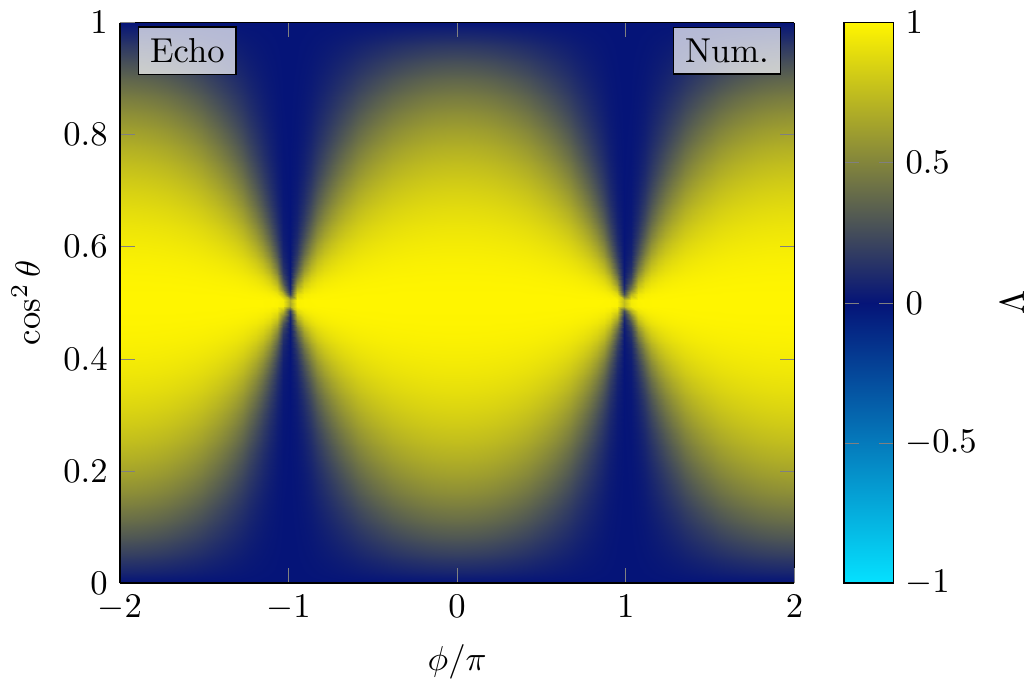}
    \end{subfigure}
    
        \begin{subfigure}
        \centering
        \includegraphics[height=5.8cm]{tikzfigures-figure2.pdf}
    \end{subfigure}
    \hfill
    \begin{subfigure}
        \centering
        \includegraphics[height=5.8cm]{tikzfigures-figure3.pdf}
    \end{subfigure}
    
    \begin{subfigure}
        \centering
        \includegraphics[height=5.8cm]{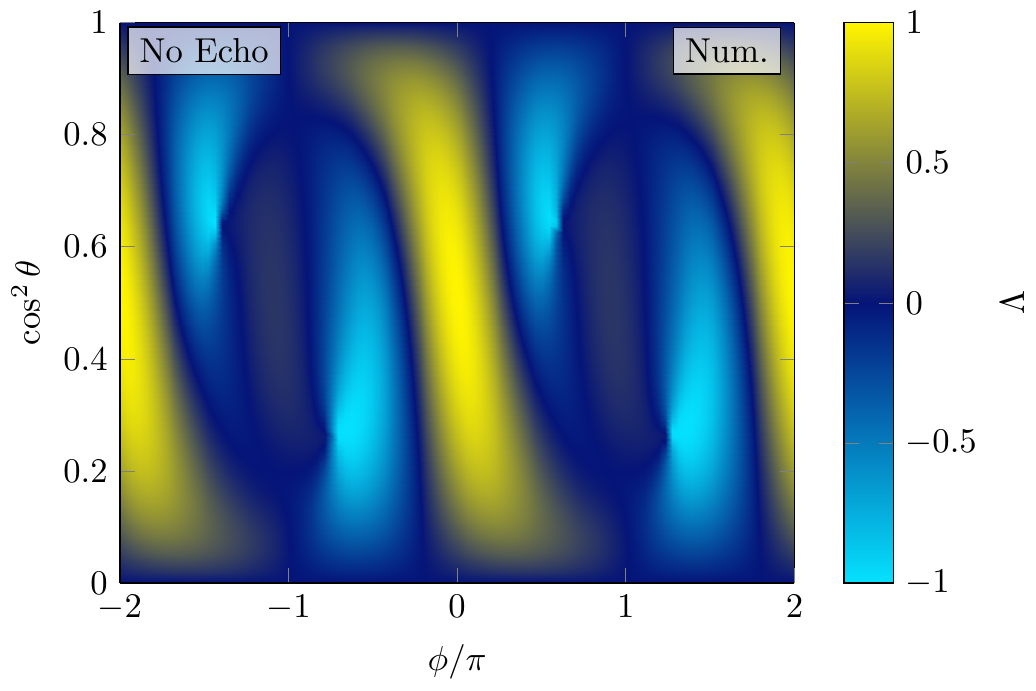}
    \end{subfigure}
    \hfill
    \begin{subfigure}
        \centering
        \includegraphics[height=5.8cm]{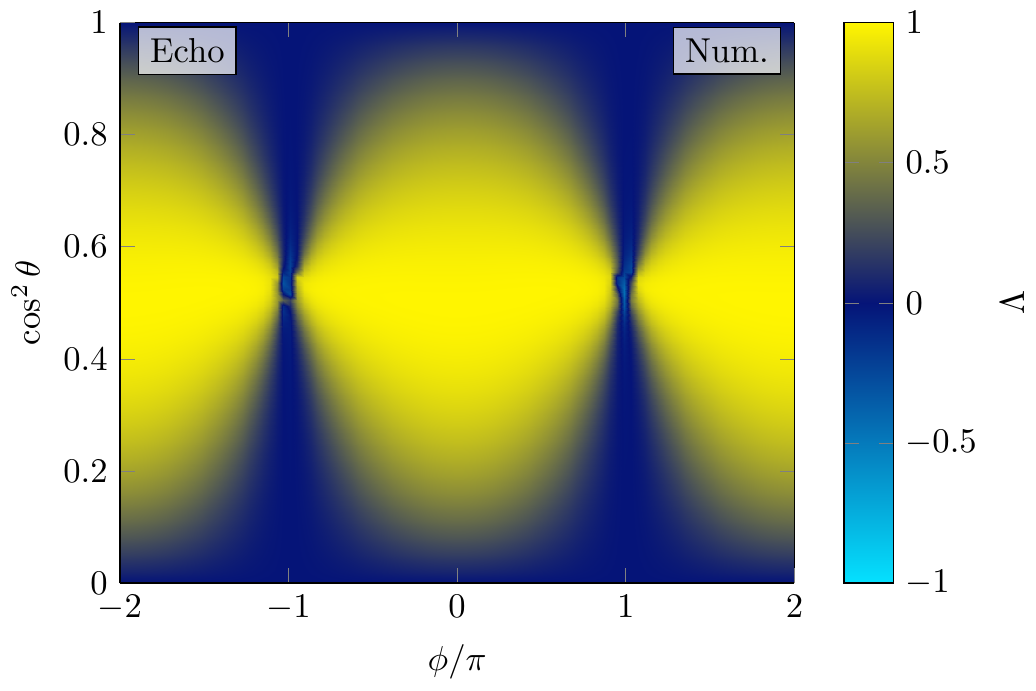}
    \end{subfigure}
    \caption{Sequence visibility $\Lambda$ obtained from numerical simulation with $ x_0=100$ and $ \eta=2$. The dimensionless expansion parameter is varied from top to bottom: $\Sigma_\eta/(Tw)=0.25, 0.5, 1$.}
    \label{fig:timescalesapp}
\end{figure*}

To continue with Eqs. \eqref{eq:cond1.1app} and \eqref{eq:cond2.1app}, we need to understand how $\dd b_{nm}^{(p)}/\dd s$ relates to $b_{nm}^{(p)}$ for $n\neq m$. For that, we restrict ourselves to the example of a two level system, Eq.\ \eqref{eq:HamMatrix}. In the following, we make an argument based on induction for the approximation
\begin{equation}\label{eq:ddscaling}
    \left| \frac{\dd}{\dd s}b_{nm}^{(p)}\right|\sim \Omega_\eta (x^2+1)^{\frac{\eta-1}{2}}\,|b_{nm}^{(p)}| \qquad n\neq m.
\end{equation}
The argument relies on the basic observation that all operators $\Delta_{10}$, $M_{10}$ and $\dd/\dd s=\dot x
\,(\partial \sqrt{x^2+1}/\partial x)\, \partial/\partial \sqrt{x^2+1}$, used to compute the coefficients $b_{nm}^{(p)}$, are polynomial in $\sqrt{x^2+1}$ with rational exponents, see Eqs. \eqref{eq:deltaspecific}, \eqref{eq:Mspecific} and \eqref{eq:dotansatz}. We begin the argument by checking that Eq.\ \eqref{eq:ddscaling} holds for the first-order coefficients found in Sec.\ \ref{sec:deriveAPT}. Taking the derivative of the only $n\neq m$, non-constant, first-order coefficient, we get
\begin{equation}\label{eq:firstorder10scaleapp}
    \left|\frac{\dd}{\dd s} b_{10}^{(1)}\right|=\Omega_\eta (x^2+1)^{\eta/2} \left|\frac{\partial \sqrt{x^2+1}}{\partial x}\right|\left|\frac{\partial (M_{10}/\Delta_{10})}{\partial \sqrt{x^2+1}}\right|.
\end{equation}
Since $\Delta_{10}$ and $M_{10}$ are polynomials in $\sqrt{x^2+1}$, we make the assertion 
\begin{align}
    \left|\frac{\partial (M_{10}/\Delta_{10})}{\partial \sqrt{x^2+1}}\right|&=|3-\eta|\left| \frac{M_{10}/\Delta_{10}}{ \sqrt{x^2+1}} \right|\\
    &\sim\left|\frac{M_{10}/\Delta_{10}}{ \sqrt{x^2+1}}\right|=\frac{|b_{10}^{(1)}|}{ \sqrt{x^2+1}}.
\end{align}
Combining this with Eq.\ \eqref{eq:firstorder10scaleapp} and dropping $|\partial\sqrt{x^2+1}/\partial x|$ as it is unimportant, we conclude that $b_{10}^{(1)}$ fulfills Eq.\ \eqref{eq:ddscaling}. To complete the induction, we show that if the coefficients of order $p$ fulfill
Eq.\ \eqref{eq:ddscaling}, then also the $p+1$ order coefficients should fulfill Eq.\ \eqref{eq:ddscaling}. We rewrite Eq.\ \eqref{eq:nneqm} using the hypothesis in Eq.\ \eqref{eq:ddscaling},
\begin{align}\nonumber
     b_{nm}^{(p+1)} &\sim \frac{i\Omega_\eta (x^2+1)^{\frac{\eta-1}{2}}}{\Delta_{nm}} b_{nm}^{(p)}(s)\\
    &+\sum_{k\neq n,m} \frac{iM_{nk}(s)}{\Delta_{nm}(s)}b_{km}^{(p)}+\frac{iM_{nm}(s)}{\Delta_{nm}(s)}b_{mm}^{(p)} \label{eq:inductivestep}
\end{align}
This equation consists of polynomials in $\sqrt{x^2+1}$ and $n\neq m$ coefficients of order $p$, which by the hypothesis fulfills Eq.\ \eqref{eq:ddscaling}. Therefore also the coefficients of order $p+1$ obeys Eq.\ \eqref{eq:ddscaling}. The only exception in Eq.\ \eqref{eq:inductivestep} is the last term with the $n=m$ coefficient. However, for large $|x|$, this coefficient is almost constant as the tails of the integrals are very close to zero and it is unimportant. For $|x|\sim 1$, all of the $p$-order coefficients are of the same magnitude, $(\Omega_\eta/w)^p$, and thus the coefficient of order $p+1$ still fulfills Eq.\ \eqref{eq:ddscaling}. This completes the argument. 

A heuristic argument that leads to the same scaling behavior for large $x$ is that whatever $\dd/\dd s=\dot x\, \dd/\dd x$ acts on, gets multiplied by $\dot x$ while a power of $x$ gets subtracted from the differentiation $\dd/\dd x$.

\begin{figure*}
    \centering
    \begin{subfigure}
        \centering
        \includegraphics[height=5.8cm]{tikzfigures-figure5.pdf}
    \end{subfigure}
    \hfill
    \begin{subfigure}
        \centering
        \includegraphics[height=5.8cm]{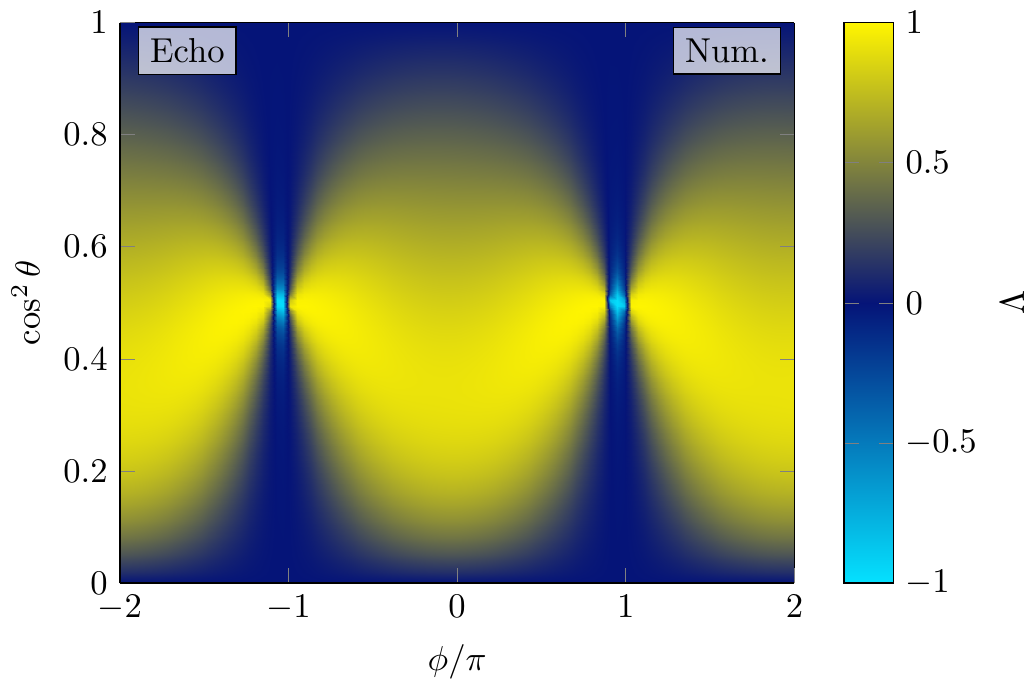}
    \end{subfigure}
    
    \begin{subfigure}
        \centering
        \includegraphics[height=5.8cm]{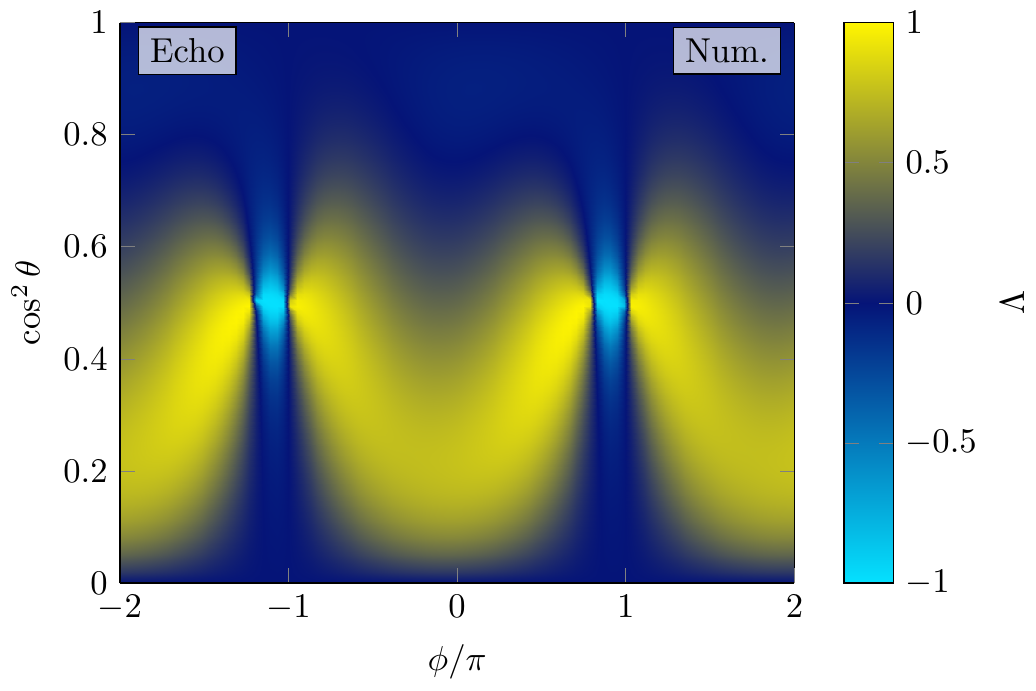}
    \end{subfigure}
    \hfill 
    \begin{subfigure}
        \centering
        \includegraphics[height=5.8cm]{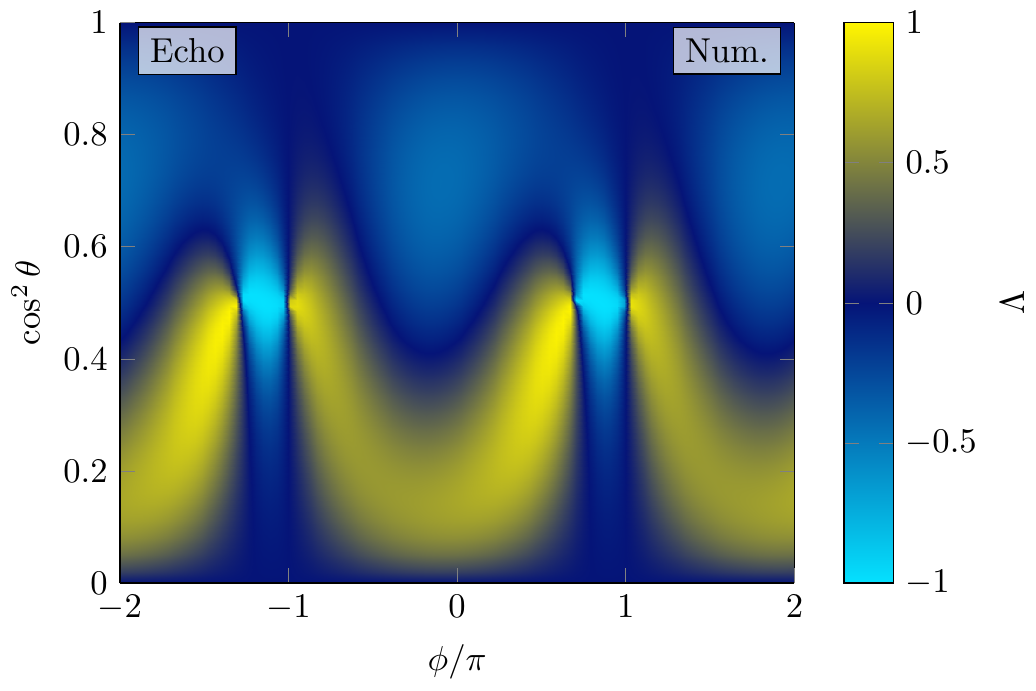}
    \end{subfigure}
    \caption{Sequence visibility $\Lambda$ obtained from numerical simulation with parameters $ x_0=100, \Sigma_\eta/(Tw)=0.5$ and $ \eta=2$ for the echo protocol. The additional phase $\delta\phi$ used is varied from top left to bottom right with increasing offset from the ideal point: $\delta\phi=2\pi, 1.05\cdot(2\pi), 1.1\cdot(2\pi),1.15\cdot(2\pi)$.}
    \label{fig:echorobustness}
\end{figure*}

\begin{figure*}
    \centering
    \begin{subfigure}
        \centering
        \includegraphics[height=5.8cm]{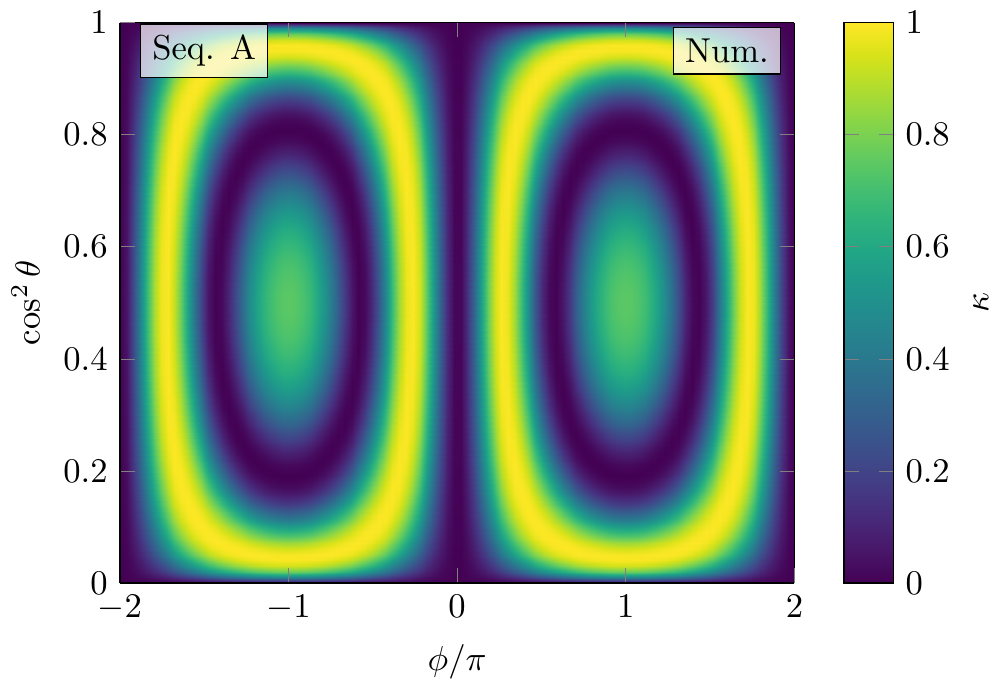}
    \end{subfigure}
    \hfill
    \begin{subfigure}
        \centering
        \includegraphics[height=5.8cm]{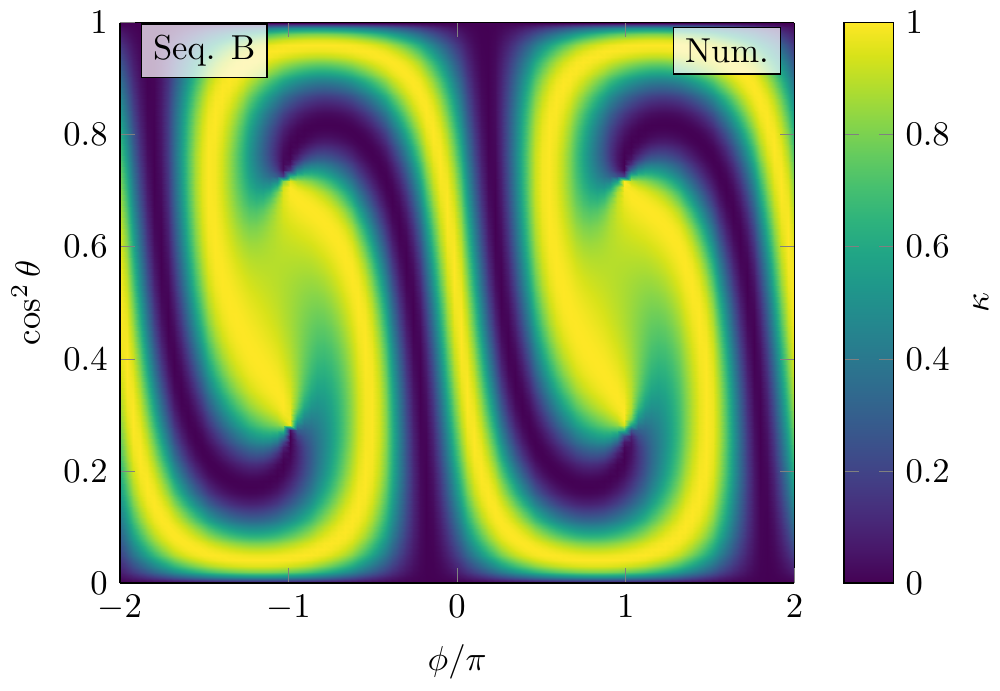}
    \end{subfigure}
    
    \begin{subfigure}
        \centering
        \includegraphics[height=5.8cm]{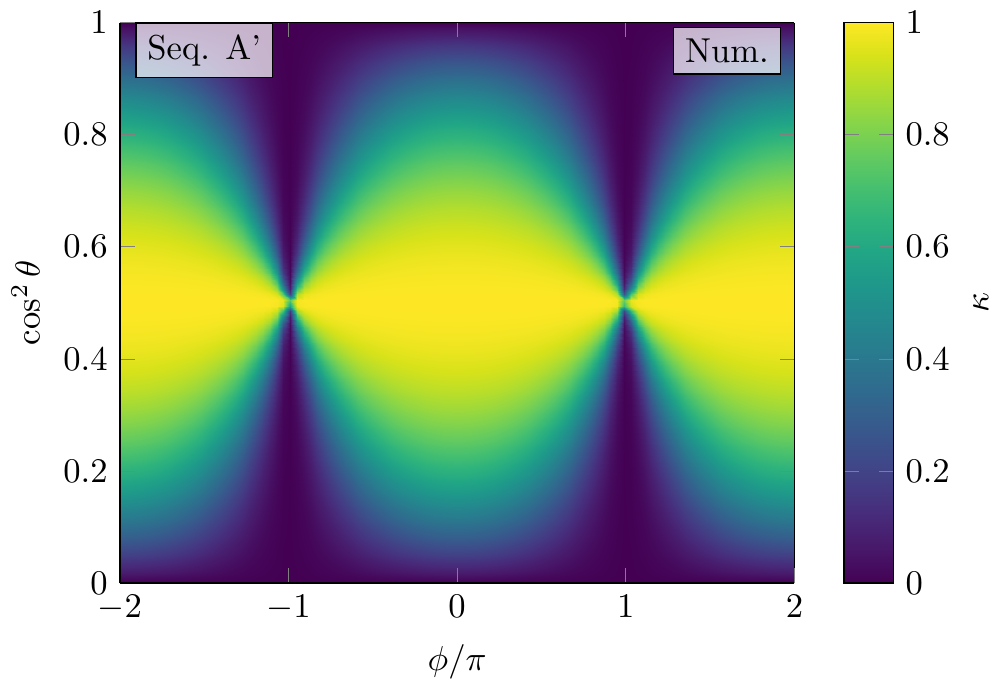}
    \end{subfigure}
    \hfill
    \begin{subfigure}
        \centering
        \includegraphics[height=5.8cm]{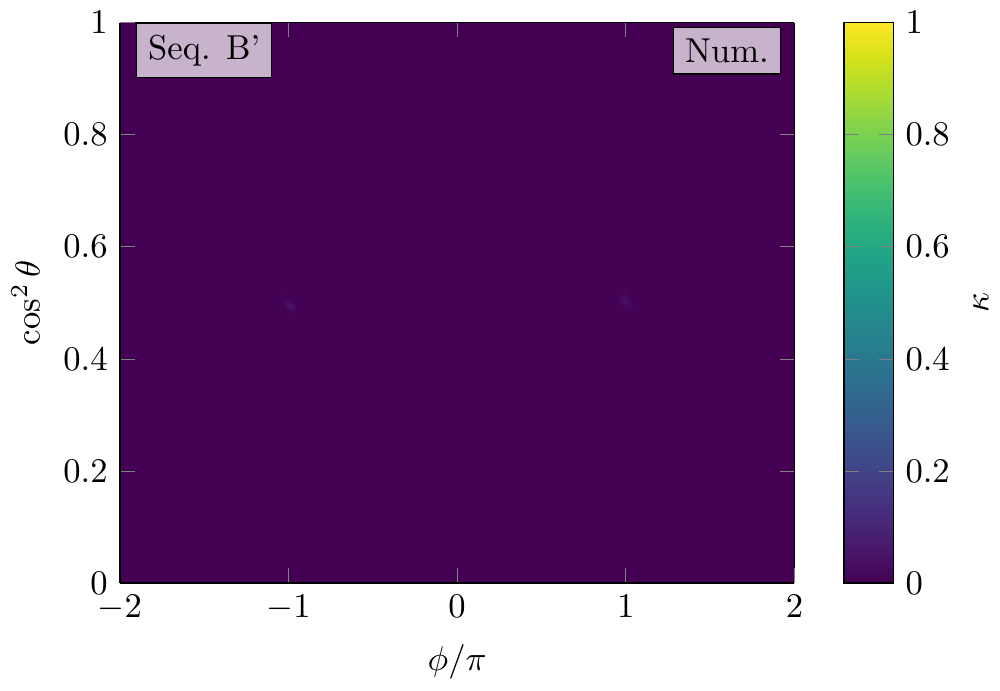}
    \end{subfigure}
    \caption{The probability of finding the state $\ket{0}_\text{M12}$ after each sequence. Numerical results with parameters  $ \eta=2, \Sigma_\eta/(Tw)=0.5$ and $x_0=100$.}
    \label{fig:sequenceprobability}
\end{figure*}

We may now use Eq.\ \eqref{eq:ddscaling} to rewrite Eqs.\ \eqref{eq:cond1.1app} and \eqref{eq:cond2.1app} and extract the corresponding adiabatic conditions,
\begin{align}\label{eq:cond1.1.1app}
    \frac{\Omega_\eta (x^2+1)^\frac{\eta-1}{2}}{T|\Delta_{nm}|}&\ll1\qquad n\neq m,\\ \label{eq:cond2.1.1app}
    \int_0^s\Omega_\eta (x^2+1)^\frac{\eta-1}{2}\frac{|M_{nm}|}{T|\Delta_{nm}|}\, \dd s'&\ll1\qquad n\neq m.
\end{align}
That is, if Eqs.\ \eqref{eq:cond1.1.1app} and \eqref{eq:cond2.1.1app} are satisfied, then also Eqs.\ \eqref{eq:cond1.1app} and \eqref{eq:cond2.1app} are satisfied. By inserting $\Delta_{10}$ and $M_{10}$ we arrive at the conditions in Eqs.\ \eqref{eq:cond1} and \eqref{eq:cond3}.

As a final remark, we discuss the APT prediction for $\eta>2$. The conditions in Eqs.\ \eqref{eq:cond1} and \eqref{eq:cond1.1.1app}, that gives the APT prediction for $\eta>2$, rely on Eq.\ \eqref{eq:ddscaling} whose proof is somewhat heuristic. The APT prediction for $\eta>2$ is therefore approximated but still required to achieve adiabaticity as shown in Fig.\ \ref{fig:transitionandphase}. The conditions in Eqs.\ \eqref{eq:cond1} and \eqref{eq:cond1.1.1app} are important to ensure that higher-order contributions in the adiabatic expansion do not grow with the order. These conditions do not appear in the first-order coefficients.
It may therefore be possible to relax the requirement in Eq.\ \eqref{eq:sumcond}, replacing the $\ll$ with $<$, while still requiring that the first-order coefficients are small. Convergence of the adiabatic expansion in Eq.\ \eqref{eq:APTansatz} is then ensured by the geometric series. This would relax the condition for adiabaticity in the region $2<\eta\leq 3$ from $\Omega_\eta/(Tw)\ll x_0^{\eta-2}$ to $\Omega_\eta/(Tw)< x_0^{\eta-2}$. The other conditions in Eqs.\ (\ref{eq:cond2}, \ref{eq:cond4}) would still be in effect.

\section{Sequence visibility at different time scales}\label{app:Visibilitytime scales}

In Fig.\ \ref{fig:timescalesapp}, we display the sequence visibility from numerical simulation for different values of the dimensionless expansion parameter. We show results for decreasing $T$ values from top to bottom. The panels in the left column show the protocol without the flux echo. Since this protocol is sensitive to the dynamical phase, we observe an increased number of fringes in the top left panel where the operation time is slower. In the bottom left panel, we see fewer fringes but also distortions due to nonadiabatic errors. In the right column, we show results for the flux echo protocol that cancels out the contribution from the dynamical phase. For this reason, we only see the contribution from the geometric phase which is insensitive to the time of operation as long as it is adiabatic.

The number of fringes $\nu$ in the left column panels can be theoretically estimated. For symmetric couplings, $\theta=\pi/4$, the sequence visibility simply becomes
\begin{align}
    \Lambda&=\cos(2\,\theta^D)\\
    &=\cos( \frac{2\sqrt{\pi}\Gamma(\frac{\eta}{2})}{\Gamma(\frac{\eta+1}{2})}\frac{Tw}{\Omega_\eta}\sin(\phi/2)).
\end{align}
The number of fringes can then be counted by the number of times $\Lambda$ is $\pm1$. In the region $-\pi<\phi<\pi$, the number of fringes is well-approximated by
\begin{equation}\label{eq:fringes}
    \nu=2 \left\lfloor \frac{2\,\Gamma(\frac{\eta}{2})}{\sqrt{\pi}\Gamma(\frac{\eta+1}{2})}\frac{Tw}{\Omega_\eta} \right\rfloor +1,
\end{equation}
for the optimal path found in this paper. Here, $\lfloor \cdot \rfloor$ is the floor function.
In agreement with the left column in Fig. \ref{fig:timescalesapp}, Eq.\ \eqref{eq:fringes} predicts $11$, $5$ and $3$ fringes for the top, middle and bottom panels.

\section{Robustness of flux echo}\label{app:RobustEcho}

In Fig.\ \ref{fig:echorobustness}, we display the sequence visibility $\Lambda$ for the echo protocol at different values of the additional SC phase $\phi\to\phi+\delta\phi$. In the top left panel, we show the ideal situation of $\delta\phi=2\pi$. In top right and bottom panels we tune slightly away from the optimal point ($\delta\phi=2\pi$) by $5\%$, $10 \%$ and $15\%$. A $5\%$ offset, as shown in the top right panel, still results in a large region in parameter space with good visibility. At a $10\%$ offset, as shown in the bottom left panel, the region size and visibility is slightly reduced and shifted to nonzero coupling asymmetry. However, even for $10\%$ error in $\delta\phi$, a high visibility can be reached by tuning $\theta$, which gives the ratio between $w_3$ and $w_4$. At $15\%$ offset, as shown in the bottom right panel, the dynamical phase plays a significant role and reduces the visibility. 

\section{Measurement signature for each sequence}\label{app:SignatureSequence}

In Fig.\ \ref{fig:sequenceprobability}, we resolve the sequence visibility into the specific probabilities after each sequence. We display the probability $\kappa$ to end up in the $\ket{0}_\text{M12}$ state. In the top panels, we show $\kappa$ for sequences A and B. Besides weak nonadiabatic corrections, sequence A only gets contributions from the dynamical phase and sequence B gets contributions from both geometric and dynamical phases. For sequences A' and B', where the flux echo is in effect, there is no contribution from the dynamical phase. In this case, only sequence A' gets a contribution from the geometric phase, this is the reason why $\kappa$ remains zero after sequence B'.

\bibliographystyle{apsrev4-2}
\bibliography{Majorana}

\end{document}